\documentclass[journal]{IEEEtran}

\IEEEoverridecommandlockouts
\usepackage{algorithm}
\usepackage{algpseudocode}
\usepackage{amsmath,amssymb,amsfonts}

\usepackage{textcomp}
\usepackage{xcolor}

\ifCLASSOPTIONcompsoc
\usepackage[caption=false,font=normalsize,labelfont=sf,textfont=sf]{subfig}
\else
\usepackage[caption=false,font=footnotesize]{subfig}
\fi

\usepackage{graphicx}
\usepackage{color}
\usepackage{xfrac}
\usepackage{amsthm}
\usepackage{relsize}
\usepackage{enumerate}
\usepackage{booktabs}
\usepackage{morefloats}
\usepackage{cite}
\usepackage{bm}
\usepackage{stfloats}
\usepackage{setspace}
\usepackage{url}
\usepackage{ulem}
\usepackage{comment}
\usepackage[figuresright]{rotating}
\usepackage{multirow}
\usepackage{lscape}
\usepackage{adjustbox}
\usepackage{bm}
\usepackage{tablefootnote}
\usepackage{balance}

\usepackage{xr}

\newcounter{mytempeqncnt}

\begin{document}

\title{To Sense or Not To Sense: A Delay Perspective}
\author{
    Xinran Zhao and Lin Dai
    \vspace{-0.6cm}
    \thanks{This paper will be presented in part at the IEEE International Conference on Communications, Denver, USA, June 2024.}
    \thanks{The authors are with the Department of Electrical Engineering, City University of Hong Kong, Hong Kong (e-mail: xrzhao3-c@my.cityu.edu.hk; lindai@cityu.edu.hk).}
}
\maketitle

\begin{abstract}

With the ever-growing demand for low-latency services in machine-to-machine (M2M) communications, the delay performance of random access networks has become a primary concern, which critically depends on the sensing capability of nodes. To understand the effect of sensing on the optimal delay performance, the challenge lies in unifying the delay analysis of sensing-free Aloha and sensing-based Carrier Sense Multiple Access (CSMA) with various design features such as backoff and connection-free or connection-based. In this paper, based on a unified analytical framework, the mean queueing delay of data packets with Aloha and CSMA is characterized and optimized, with which the upper-bound of sensing time for CSMA to outperform Aloha in terms of the minimum mean queueing delay is further obtained. The analysis is also applied to the Random Access-Based Small Data Transmission (RA-SDT) schemes in 5G networks to investigate when and how significant their delay performance can be improved by sensing, which sheds important insights into practical access protocol design.

\end{abstract}


\section{Introduction}\label{sec1}

With the explosive growth of machine-type devices (MTDs), wireless communication networks are undergoing a paradigm shift from human-to-human (H2H) communications toward machine-to-machine (M2M) communications. The ubiquitous deployment of MTDs has catalyzed a wide range of new applications in various fields, such as industry automation, intelligent transportation and healthcare, many of which are mission critical requiring the end-to-end latency in the order of milliseconds or even sub-milliseconds \cite{3gpp.22.104,8644245}. Providing such low-latency services for a massive number of MTDs has posed an unprecedented challenge for the design of next-generation wireless networks \cite{8869705}.

In enabling the massive access of MTDs for M2M communications, random access has played a pivotal role \cite{leyva2019random}. With random access, nodes independently decide when to transmit, which avoids heavy scheduling overhead thanks to its distributed and low-cost nature. Due to the lack of coordination among nodes, nevertheless, concurrent transmissions may collide, leading to transmission failures and causing excessively long delay in retransmissions if the access parameters are not properly selected. To reduce the collisions, Carrier Sense Multiple Access (CSMA) \cite{kleinrock1975packet} has been widely adopted, with which nodes first sense the channel before their transmissions, and transmit only if the channel is sensed idle. Despite alleviated contention among nodes, sensing may not always bring improvements in the delay performance compared to sensing-free random access (referred to as Aloha \cite{abramson1970aloha}).\footnote{Both sensing-free random access and sensing-based random access have been extensively applied in practical systems. The random access procedure of 5G networks in the licensed spectrum \cite{3gpp.38.321}, for instance, is a sensing-free random access protocol. On the other hand, examples of sensing-based random access include the Distributed Coordination Function (DCF) in WiFi networks \cite{ieee_802_11_MAC} and the random access procedure for 5G networks in the unlicensed spectrum \cite{3gpp.37.213}.} Due to the propagation delay, each node needs to continuously sense for a while to ensure that the channel is indeed idle. Apparently, whether the benefit of sensing outweighs the cost is closely determined by the sensing time.

Intuitively, there exists an upper-bound of sensing time, only below which sensing is beneficial. Such an upper-bound in terms of the maximum network throughput has been obtained with the collision receiver in \cite{6525472} and the capture receiver in \cite{8779696}, respectively. The analysis was further extended in \cite{8675765} from connection-free random access to connection-based random access,\footnote{In practice, examples of connection-free random access include the 2-step random access procedure of 5G networks in the licensed spectrum \cite{3gpp.38.321} and the basic access mechanism of DCF in WiFi networks \cite{ieee_802_11_MAC}. Examples of connection-based random access include the 4-step random access procedure of 5G networks in the licensed spectrum \cite{3gpp.38.321} and the request-to-send/clear-to-send (RTS/CTS) access mechanism of DCF in WiFi networks \cite{ieee_802_11_MAC}.} where a request is sent to the receiver to establish a connection before each data packet transmission. Note that to maximize the network throughput, nodes' queues should be pushed to saturation, i.e, the queues are always busy, with which the queueing delay of data packets grows with time unboundedly. As a result, the analysis in \cite{6525472,8779696,8675765} sheds little light on when sensing is beneficial to the delay performance, which is of great interest considering the growing demand for supporting low-latency services in M2M communications.

For characterization of the criteria of beneficial sensing on delay, the delay performance of sensing-free and sensing-based random access schemes with various design features should be analyzed, optimized and compared based on a unified framework, which unfortunately has remained elusive. To understand the challenges of unifying the delay analysis of random access networks, let us first present an overview of the previous studies below.

\subsection{Delay Analysis of Random Access Networks}\label{sec1-1}

With each node equipped of a buffer to store incoming data packets, essentially a random access network can be regarded as a multi-queue-single-server system. The queueing (end-to-end) delay of each data packet consists of two parts: \romannumeral1) the waiting time in the queue, i.e., the time from arrival till being the Head-of-Line (HOL) packet, and \romannumeral2) the service time or access delay, i.e., the time from being the HOL packet till being successfully transmitted.

\subsubsection{Access Delay}\label{sec1-1-1}

The access delay mainly comes from nodes' contention process. Thus, many studies put their focus on the characterization of contention among nodes and excluded nodes' queues from the models. Specifically, early studies often assumed that each node is equipped with a one-packet buffer, and focused on the state characterization of the total number of backlogged nodes (i.e., nodes with nonempty buffer) for Aloha \cite{1092814,1092823} and CSMA \cite{tobagi1980performance,1095881}. Capturing the essence of contention among nodes, such an approach greatly simplifies the access delay analysis, and has a huge impact on later studies of throughput and access delay optimization of various Aloha-based \cite{6824261,7447749,9001113,9785915,10025558} and CSMA-based \cite{6510019,6384611,10075522} networks. Yet with nodes' queues ignored, the adaptive tuning of access parameters is usually performed based on the time-varying network status such as the total number of backlogged nodes in the network, which is difficult to track and estimate accurately in practice.

In addition to the one-packet-buffer model, another simplification is to only consider the saturated condition, where each node always has data packets in its buffer. For instance, in \cite{6751602} for Aloha and \cite{840210,1512134} for CSMA, the focus was placed on the state characterization of nodes' backoff process. Despite accurate evaluation of the throughput and mean access delay, the mean queueing delay of data packets is infinite when nodes' queues are saturated.

\subsubsection{Queueing Delay}\label{sec1-1-2}

With nodes' queues included, various models have been proposed to analyze the queueing delay performance \cite{1095740,hofri1983analysis,1096221,1102713,6373677,10319458,sheikh2004performance,1096647,cantieni2005performance,4100720,5487530}. For a two-node connection-free Aloha network, the mean queueing delay of data packets with Bernoulli arrivals was first obtained in \cite{1095740} by deriving the joint distribution of nodes' queue lengths. The analysis was further extended to a two-node connection-based Aloha network and a two-node CSMA network in \cite{hofri1983analysis} and \cite{1096221}, respectively. It is, however, difficult to obtain the joint queue length distribution of nodes when the network size exceeds two. Therefore, most studies have considered the symmetric scenario where nodes have identical input rates and access parameters. For instance, in \cite{1102713,6373677,10319458} for connection-free Aloha, \cite{sheikh2004performance} for connection-based Aloha and \cite{1096647,cantieni2005performance,4100720,5487530} for CSMA, various Markov chains were established to either jointly model the queue length of a tagged node and the total number of backlogged nodes \cite{1102713,1096647}, or characterize the state transition of a tagged node \cite{6373677,10319458,sheikh2004performance,cantieni2005performance,4100720,5487530}. To calculate the mean queueing delay, nevertheless, iterative algorithms were often developed for solving a set of nonlinear equations, leading to high computational complexity and few insights in optimal tuning of access parameters for delay optimization.

In fact, for each node's queue, given the arrival process of packets, the key to deriving the mean queueing delay lies in the characterization of its service process, which is determined by the aggregate activities of nodes' HOL packets. It was demonstrated in \cite{6205590} that for connection-free Aloha, by properly modeling the behavior of HOL packets in each node's queue, the first and second moments of service time can be derived as explicit functions of transmission probability of each node, which further facilitates the delay optimization. Such a HOL-packet modeling approach has been adopted to characterize and optimize the mean access delay for connection-free Aloha in \cite{6205590,9429212} and CSMA in \cite{6525472}, and the mean queueing delay for connection-based Aloha in \cite{10154598}.

Note that in the above studies, different random access schemes were usually considered in a separate manner. To unify the delay analysis of random access networks, the challenge lies in establishing a unified analytical framework that can incorporate various features of random access, but still yield simple and accurate characterization of each node's service time distribution. As we will demonstrate in this paper, based on a general Markov renewal process of HOL packets, where the states, state transition probabilities, and holding time at each state are determined by design features, a unified analytical framework for random access can be constructed, which enables not only the delay optimization for a given random access scheme, but also a comprehensive comparison of optimal delay performance across different random access schemes.
\vspace{-0.1cm}
\subsection{Our Contributions}\label{sec1-2}

In this paper, a comparative study of the optimal delay performance of sensing-free Aloha and sensing-based CSMA is conducted to characterize the criteria for beneficial sensing. Specifically, a unified analytical model for random access is proposed to incorporate various design features, including sensing-free or sensing-based, connection-free or connection-based, and backoff,\footnote{In random access networks, backoff schemes are often adopted to resolve transmission failures. Specifically, each node can adjust its transmission probability according to the number of transmission failures experienced by its HOL packet. In practice, Constant Backoff and Binary Exponential Backoff are two widely adopted backoff schemes in cellular networks and WiFi networks.} with which the mean queueing delay of data packets with Bernoulli arrivals is derived as an explicit function of the initial transmission probability of each node and further minimized by optimally choosing the initial transmission probability.

Based on the delay analysis, the upper-bound of sensing time for CSMA to outperform Aloha in terms of the minimum mean queueing delay is further characterized. Different from the throughput-optimal sensing bound $\sigma_C^{\ast}$ derived in \cite{6525472,8779696,8675765} that is independent of the backoff scheme and the data input rates of nodes, the delay-optimal sensing bound $\tilde{\sigma}_C^{\ast}$ is found to be critically dependent on them. It is shown that $\tilde{\sigma}_C^{\ast}$ could be much higher than $\sigma_C^{\ast}$, which indicates that a more substantial improvement can be achieved in terms of the optimal delay performance than the optimal throughput performance by the use of sensing. To further demonstrate the practical implications of the analysis, a case study on the Random Access-Based Small Data Transmission (RA-SDT) schemes of 5G networks is conducted. The analysis suggests that incorporating sensing into the existing sensing-free RA-SDT may significantly improve the delay performance, especially for the grant-free 2-step RA-SDT.

The remainder of the paper is organized as follows. Section \ref{sec2} presents the system model. A unified analytical framework is established in Section \ref{sec3}, based on which a unified delay analysis is presented in Section \ref{sec4}. The delay-optimal sensing bound is characterized and further applied to 5G networks in Section \ref{sec5}. Finally, conclusions are summarized in Section \ref{sec6}.

For superscripts and subscripts throughout the paper, $A$ and $C$ denote Aloha and CSMA, respectively. $P$ and $N$ denote connection-free and connection-based, respectively.

\section{System Model}\label{sec2}

\begin{figure}[t]
    \centering
    \centering
    \includegraphics[width=0.7\linewidth, height=0.22\linewidth]{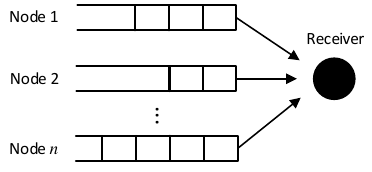}
    \caption{Graphic illustration of an $n$-node buffered random access network.}
    \label{fig2-1}
\end{figure}

Consider a slotted random access network where $n$ nodes transmit to a single receiver. Assume that each node has an input bitstream with rate $\lambda_b$ bit/s/Hz, which is divided into data packets with encoding rate $R$ bit/s/Hz, and stored in a buffer with infinite size, as illustrated in Fig. \ref{fig2-1}. The time axis is divided into multiple slots with equal length $\sigma$ milliseconds (ms). All the nodes are synchronized and can transmit only at the beginning of a time slot. Assume that the transmission of a data packet lasts for $L$ ms. The packet input rate $\lambda$ of each node's queue, i.e., the average number of arrival data packets per time slot, can then be written as
\begin{equation}
    \lambda = \tfrac{\lambda_b\sigma}{RL}.
\label{eq2-1}
\end{equation}
For the sake of clarity, in this paper, assume that the arrivals of data packets of each node's queue follow a Bernoulli process with rate $\lambda$.

Due to incoordination among nodes, concurrent transmissions may lead to failures. Assume the collision model at the receiver, that is, if more than one node transmits simultaneously, a collision occurs and all of them fail. A transmission is successful if and only if there are no concurrent transmissions, in which case the receiver would broadcast an acknowledgement (ACK) to nodes. If no ACK is received within a certain time, nodes infer that the preceding transmission has failed, and retransmit with a certain probability. Assume that the transmission probability of each node can be adjusted according to the number of transmission failures experienced by its HOL packet. Specifically, the transmission probability after the $k$-th failure can be expressed as $q_0\mathcal{Q}(k)$, $k=0,1,\cdots$, where $q_0$ is the initial transmission probability, and $\mathcal{Q}(k)$ is the backoff function and usually chosen as a monotonic non-increasing function of $k$. In general, we have $\mathcal{Q}(0)=1$, $\mathcal{Q}(k+1)\leq \mathcal{Q}(k)$ for $k=0,1,\cdots,K-1$, and $\mathcal{Q}(k)=\mathcal{Q}(K)$ for $k\geq K$, where $K$ is the cutoff phase and usually adopted to prevent the transmission probability from being excessively small.

\vspace{-0.2cm}
\subsection{Minimum Mean Queueing Delay of Data Packets}\label{sec2-1}

In this paper, we focus on the delay performance of random access networks. For each data packet, define its queueing delay $T$ as the number of time slots from its arrival till its successful transmission. The queueing delay performance closely depends on whether the queue is saturated or not. Let $\mu$ denote the service rate of each node's queue (i.e., the average number of data packets that can be successfully transmitted per time slot given that the queue is nonempty). If the packet input rate $\lambda<\mu$, the queue is unsaturated with a non-zero probability of being empty. Otherwise, the queue is saturated, and the queueing delay of data packets grows with time unboundedly.

It can be seen that the mean queueing delay of data packets $\overline{T}$ crucially depends on the service rate of each node's queue $\mu$, which is further determined by each node's access parameters, e.g., the initial transmission probability $q_0$. For finite $\overline{T}$, $q_0$ should be properly selected to ensure that each node's queue is unsaturated. Define the unsaturated region of initial transmission probability $q_0$ as
\begin{equation}
    S_{U^n}(q_0,\hat{\lambda}) = \{q_0:\hat{\lambda}<\hat{\mu}\},
\label{eq2-3}
\end{equation}
where $\hat{\lambda}=n\lambda$ and $\hat{\mu}=n\mu$ are the aggregate packet input rate and aggregate packet service rate, respectively. Note that if $\hat{\lambda}$ is too large, each node's queue would always be saturated no matter what value the initial transmission probability $q_0$ is set to. To be more specific, $S_{U^n}(q_0,\hat{\lambda})$ is nonempty only when $\hat{\lambda}<\hat{\lambda}_{\max}$, where $\hat{\lambda}_{\max}$ is defined as the maximum data throughput, i.e., the maximum average number of data packets successfully transmitted per time slot. The maximum data throughput in the unit of bit/s/Hz, denoted as $\tilde{\lambda}_{\max}$, can be written from (\ref{eq2-1}) as
\begin{equation}
    \tilde{\lambda}_{\max}=\tfrac{RL}{\sigma}\cdot\hat{\lambda}_{\max}.
\label{eq2-3.5}
\end{equation}

With $q_0\in S_{U^n}(q_0,\hat{\lambda})$, the mean queueing delay of data packets $\overline{T}$ is crucially dependent on the initial transmission probability $q_0$, and can be minimized by optimally choosing $q_0$. Define the minimum mean queueing delay of data packets as
\begin{equation}
    \overline{T}_{\min} = \min_{q_0\in S_{U^n}(q_0,\hat{\lambda})} \overline{T}.
\label{eq2-4}
\end{equation}
In Section \ref{sec4}, we will derive the minimum mean queueing delay of data packets $\overline{T}_{\min}$ and the corresponding optimal initial transmission probability $q_0^{\ast}$.
\vspace{-0.1cm}
\subsection{Random Access: Sensing-Free versus Sensing-Based and Connection-Free versus Connection-Based}\label{sec2-2}

According to whether nodes have the sensing capability, random access protocols can be divided into two categories: sensing-free Aloha and sensing-based CSMA. Different from Aloha where sensing is not required, with CSMA, if a node has packets to transmit, it needs to sense the channel first. Only when the channel is sensed idle can it transmit with a certain probability at the next time slot.

\begin{figure}[t]
    \centering
    \subfloat[]{
        \includegraphics[width=0.43\textwidth, height=0.04\textwidth]{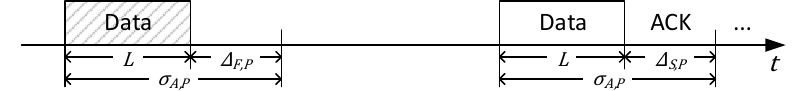}
        \label{fig2-2-1}
    }

    \subfloat[]{
        \includegraphics[width=0.43\textwidth, height=0.04\textwidth]{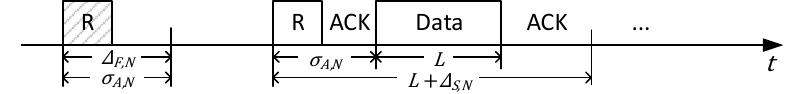}
        \label{fig2-2-2}
    }
    \caption{Graphic illustration of (a) connection-free Aloha and (b) connection-based Aloha.}
    \label{fig2-2}
\end{figure}

\begin{figure}[t]
    \centering
    \subfloat[]{
        \includegraphics[width=0.4\textwidth, height=0.04\textwidth]{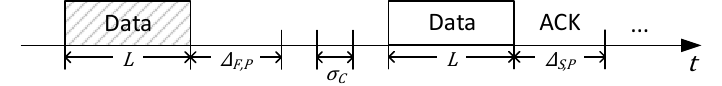}
        \label{fig2-3-1}
    }

    \subfloat[]{
        \includegraphics[width=0.4\textwidth, height=0.04\textwidth]{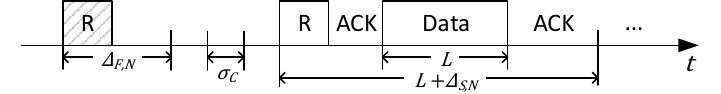}
        \label{fig2-3-2}
    }
    \caption{Graphic illustration of (a) connection-free CSMA and (b) connection-based CSMA.}
    \label{fig2-3}
\end{figure}

In addition to sensing-free or sensing-based, another important design feature of random access is connection-free or connection-based. Different from connection-free random access where every data packet has to contend, with connection-based random access, each node first sends a request, and transmits its data packet only after receiving the ACK from the receiver that indicates the successful detection of its request, i.e., a connection is established. Let $\Delta_S$ and $\Delta_F$ denote the overhead time for each successful and failed transmission, respectively, both in the unit of ms. With connection-free random access, each failed transmission and each successful transmission take $L+\Delta_{F,P}$ ms and $L+\Delta_{S,P}$ ms, respectively, as illustrated in Fig. \ref{fig2-2-1} and Fig. \ref{fig2-3-1}. With connection-based random access, each failed transmission and each successful transmission take $\Delta_{F,N}$ ms and $L+\Delta_{S,N}$ ms, respectively, as illustrated in Fig. \ref{fig2-2-2} and Fig. \ref{fig2-3-2}.

For connection-free and connection-based Aloha, the time slot length $\sigma_{A,P}$ and $\sigma_{A,N}$ are determined by the time of each data packet transmission and each request transmission, respectively, i.e.,
\begin{equation}
    \sigma_{A,P} = L+\Delta_{F,P} = L+\Delta_{S,P},\,\,\,\sigma_{A,N} = \Delta_{F,N}.
    \label{eq2-2-1}
\end{equation}
Note that with connection-based Aloha, the request process takes one time slot. Therefore, if a node successfully transmits its request, the channel would be reserved in the following $\tfrac{L+\Delta_{S,N}}{\sigma_{A,N}}-1$ time slots and not accessible to the other nodes. For CSMA, on the other hand, the time slot length $\sigma_C$ is determined by the sensing time, which is typically smaller than the time of each failed or successful transmission.

\subsection{Upper-Bound of Sensing Time for Beneficial Sensing}\label{sec2-3}

Intuitively, by introducing sensing, the collision probability can be greatly reduced. Therefore, if the sensing time $\sigma_C$ is negligible compared to the time of each packet or request transmission, CSMA would outperform Aloha. If $\sigma_C$ is comparable to the transmission time, however, the benefit from sensing may not outweigh the cost of additional time for sensing before each transmission. It can be expected that there exists an upper-bound of sensing time, only below which sensing is beneficial.

It was shown in \cite{8675765} that CSMA outperforms Aloha in terms of the maximum data throughput $\tilde{\lambda}_{\max}$ only if the sensing time $\sigma_C$ is below the throughput-optimal sensing bound $\sigma_C^{\ast}$, which depends on the data packet transmission time, overhead parameters, and whether a connection is established before each data transmission. Specifically, for connection-free and connection-based random access, $\sigma_C^{\ast,P}$ and $\sigma_C^{\ast,N}$ are given by
\begin{equation}
    \begin{split}
        \sigma_C^{\ast,P} = & e^{\tfrac{(1-e)(L+\Delta_{S,P})}{eL+\Delta_{F,P}+(e-1)\Delta_{S,P}}}\cdot(eL+\Delta_{F,P}+(e-1)\Delta_{S,P}) \\
        & - L - \Delta_{F,P},
    \end{split}
    \label{eq2-5}
\end{equation}
and
\begin{equation}
    \sigma_C^{\ast,N} = (e^{e^{-1}}-1)\Delta_{F,N},
\label{eq2-6}
\end{equation}
respectively.

In this paper, we are interested in the optimal delay performance of Aloha and CSMA. Note that the time slot length $\sigma_A$ with Aloha and $\sigma_C$ with CSMA are usually unequal. Therefore, to compare the delay performance of Aloha and CSMA, let
\begin{equation}
    \tilde{T}=\sigma\cdot\overline{T}
\label{eq2-7}
\end{equation}
denote the mean queueing delay of data packets in the unit of ms. The minimum mean queueing delay of data packets in the unit of ms, which is defined as $\tilde{T}_{\min} = \min_{q_0\in S_{U^n}(q_0,\hat{\lambda})} \tilde{T}$, can be written as $\tilde{T}_{\min}=\sigma\cdot\overline{T}_{\min}$ by combining (\ref{eq2-4}) with (\ref{eq2-7}) and noting that the time slot length $\sigma$ is independent of $q_0$. We then define the delay-optimal sensing bound $\tilde{\sigma}_C^{\ast}$ as
\begin{equation}
    \tilde{\sigma}_C^{\ast} = \max\{\sigma_C:\tilde{T}_{\min}^C\leq\tilde{T}_{\min}^A<\infty\}.
\label{eq2-8}
\end{equation}
As we will show in Section \ref{sec5}, in addition to the data packet transmission time, overhead parameters, and which type of random access is adopted, connection-free or connection-based, the delay-optimal sensing bound $\tilde{\sigma}_C^{\ast}$ further depends on the aggregate bit input rate $\tilde{\lambda}=n\lambda_b$.

\section{HOL-Packet Modeling}\label{sec3}

For a multi-queue-single-server system, the performance of nodes' queues is determined by the aggregate activities of their HOL packets. In this section, a HOL-packet model for performance analysis of random access networks will be developed by characterizing the behavior of HOL packets in a Markov renewal process and establishing the fixed-point equations of the steady-state probability of successful transmission of HOL packets.

\subsection{State Characterization of HOL Packets}\label{sec3-1}

For each node, a discrete-time Markov renewal process $(\mathbf{X}, \mathbf{V})=\{(X_j, V_j),\;j=0,1,...\}$ can be established to model the behavior of its HOL packets, where $X_j$ denotes the state of a tagged HOL packet at the $j$-th transition and $V_j$ denotes the epoch at which the $j$-th transition occurs. Let $\mathbf{X}=\{X_j\}$ denote the embedded Markov chain of $(\mathbf{X}, \mathbf{V})$ with state space $\mathcal{S}$ and steady-state probability distribution $\{\pi_u\}_{u\in\mathcal{S}}$. Let $Y_u$ denote the holding time at State $u$ in the unit of time slots, and $\tau_u$ denote the mean of $Y_u$, $u\in\mathcal{S}$. The limiting state probabilities of the Markov renewal process $(\mathbf{X}, \mathbf{V})$ are given by
\begin{equation}
    \tilde{\pi}_u = \frac{\pi_u\tau_u}{\Sigma_{v\in\mathcal{S}}\pi_v\tau_v},\,\,\,u\in\mathcal{S}.
\label{eq3-1-1}
\end{equation}

For each HOL packet, let State T denote the successful-transmission state, only at which its successful transmission occurs. The service rate of each node's queue can then be written as
\begin{equation}
    \mu = \frac{\tilde{\pi}_T}{\tau_T},
    \label{eq3-1-2}
\end{equation}
where the mean holding time at State T $\tau_T$ in the unit of time slots is given by
\begin{equation}
    \tau_T = \frac{L+\Delta_S}{\sigma}.
    \label{eq3-1-3}
\end{equation}

It is clear from (\ref{eq3-1-1})-(\ref{eq3-1-3}) that to characterize the service rate of each node's queue $\mu$, we need to determine the state transition process of HOL packets and the holding time in each state, which depend on the access mechanism, e.g., sensing-free or sensing-based. In the following subsections, for Aloha and CSMA, we will demonstrate how to establish the Markov renewal process of HOL packets and derive the service rate of each node's queue.

\subsubsection{Aloha}\label{sec3-1-1}

\begin{figure}[t]
    \centering
    \includegraphics[width=0.49\textwidth]{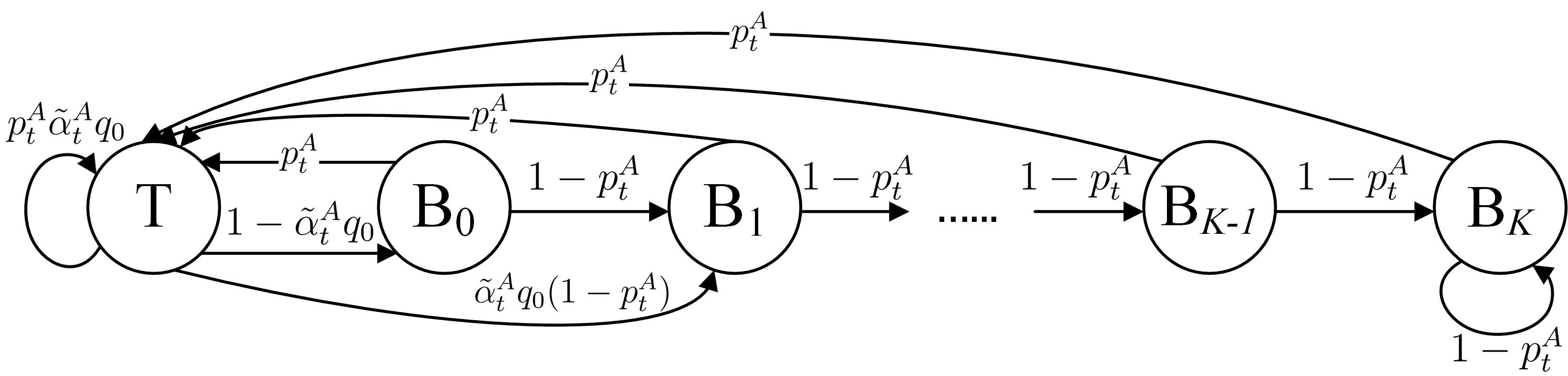}
    \caption{Embedded Markov chain $\mathbf{X^A}$ of the state transition process of an individual HOL packet in the Aloha network.}
    \label{fig3-1}
\end{figure}

Fig. \ref{fig3-1} illustrates the embedded Markov chain of state transition process of each HOL packet in the Aloha network $\mathbf{X^A}$. Each HOL packet has \romannumeral1) a successful-transmission state, State T, and \romannumeral2) $K+1$ failed-transmission/backoff states, B$_0,\cdots$, B$_K$. A State-T HOL packet would stay in State T for $\tau_T^A$ time slots if it requests to transmit and its request is successful. Otherwise, it would move to State B$_0$ if its request is suspended, or State B$_1$ if its request is failed. For a State-B$_k$ HOL packet, $k\in\{0,\cdots,K\}$, if its request of transmission is successful, it shifts to State T; otherwise, it moves to State B$_{\min\{k+1,K\}}$.

For each node, let $\tilde{\alpha}_t^A$ denote the probability that the channel is accessible at time slot $t$ given that it intends to request (i.e., its HOL packet is at State B$_k$, $k\in\{0,\cdots,K\}$, or the first time slot of State T) at time slot $t$. Note that with connection-free Aloha, $\tilde{\alpha}_t^{A,P} = 1$ as the channel is always accessible; with connection-based Aloha, $\tilde{\alpha}_t^{A,N} < 1$ as the channel is not accessible if it is reserved for data transmission. On the other hand, for each node, let $p_t^A$ denote the probability that its request of transmission is successful at time slot $t$ given that the channel is accessible and it requests at time slot $t$. The steady-state probability distribution of the embedded Markov chain $\mathbf{X^A}$ in Fig. \ref{fig3-1} can then be obtained as
\begin{equation}
    \pi_T^A \!=\! \tfrac{p^A}{p^A(1-\tilde{\alpha}^A q_0)+1},\,\pi_{B_k}^A\!=\!
    \begin{cases}
        \!(1-\tilde{\alpha}^A q_0)\pi_T^A & k\!=\!0 \\
        \!(1-p^A)^k\pi_T^A & k\!\in\!\{\!1,\!\cdots,K\!-\!1\!\} \\
        \!\tfrac{(1-p^A)^K}{p^A}\pi_T^A & k\!=\!K,
    \end{cases}
\label{eq3-1-1-1}
\end{equation}
where $\tilde{\alpha}^A = \lim \limits_{t \to \infty} \tilde{\alpha}_t^A$ and $p^A = \lim \limits_{t \to \infty} p_t^A$.

The mean holding time at State T can be obtained by combining (\ref{eq3-1-3}) with (\ref{eq2-2-1}) as
\begin{equation}
    \tau_T^A =
    \begin{cases}
        1 & {\text{for connection-free Aloha}} \\
        \tfrac{L+\Delta_{S,N}}{\Delta_{F,N}} & {\text{for connection-based Aloha}}.
    \end{cases}
\label{eq3-1-1-1.5}
\end{equation}
The holding time $Y_{{B}_k}^A$ at State B$_k$, on the other hand, can be modeled as a Geometric random variable with probability generating function:
\begin{equation}
    G_{Y_{B_k}^A}(z) = \tfrac{\tilde{\alpha}^A q_0\mathcal{Q}(k)z}{1-(1-\tilde{\alpha}^Aq_0\mathcal{Q}(k))z},
\label{eq3-1-1-2}
\end{equation}
by noting that a State-B$_k$ HOL packet is transmitted with probability $q_0 \mathcal{Q}(k)$ at each time slot given that the channel is accessible at this time slot, $k\in\{0,\cdots,K\}$. The mean holding time at State B$_k$ can then be obtained from (\ref{eq3-1-1-2}) as
\begin{equation}
    \tau_{B_k}^A = G_{Y_{B_k}^A}^\prime(1) = \tfrac{1}{\tilde{\alpha}^Aq_0\mathcal{Q}(k)}.
\label{eq3-1-1-3}
\end{equation}

Finally, the limiting state probabilities of the Markov renewal process with Aloha $\{\tilde{\pi}_u^A\}_{u\in \mathcal{S}^A}$ can be obtained by combining (\ref{eq3-1-1}) with (\ref{eq3-1-1-1})-(\ref{eq3-1-1-1.5}), (\ref{eq3-1-1-3}), based on which the service rate of each node's queue $\mu^A$ can further be derived from (\ref{eq3-1-2}) as
\begin{equation}
    \mu^A = \tfrac{1}{\tau_T^A - 1 + \tfrac{1}{p^A\tilde{\alpha}^A q_0}\Bigl(\tfrac{(1-p^A)^K}{\mathcal{Q}(K)} + \sum_{k=0}^{K-1}\tfrac{p^A(1-p^A)^k}{\mathcal{Q}(k)}\Bigr)}.
\label{eq3-1-1-5}
\end{equation}

\subsubsection{CSMA}\label{sec3-1-2}

\begin{figure}[t]
    \centering
    \includegraphics[width=0.49\textwidth, height=0.14\textwidth]{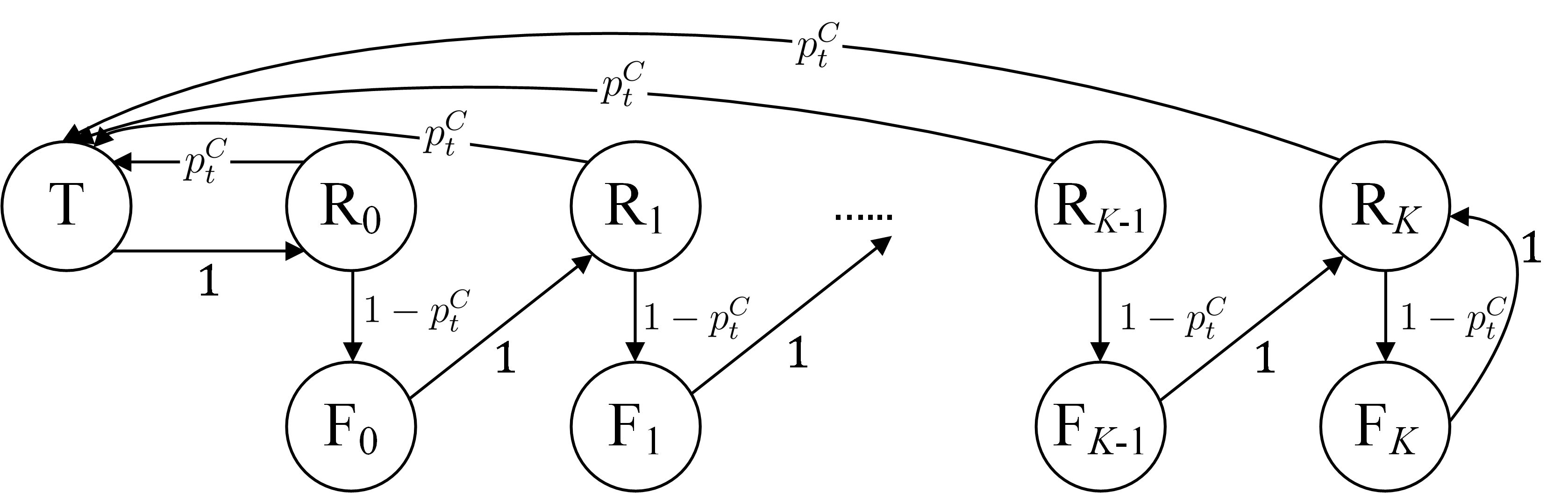}
    \caption{Embedded Markov chain $\mathbf{X^C}$ of the state transition process of an individual HOL packet in the CSMA network.}
    \label{fig3-2}
\end{figure}

With CSMA, as nodes sense the channel before each transmission, the sensing states need to be further considered and distinguished from the failed-transmission states. As Fig. \ref{fig3-2} illustrates, the states of the embedded Markov chain $\mathbf{X^C}$ of HOL packets in the CSMA network can be divided into three categories: \romannumeral1) sensing/backoff states, State R$_0,\cdots,$ R$_K$, \romannumeral2) failed-transmission states, State F$_0,\cdots,$ F$_K$, and \romannumeral3) successful-transmission state, State T. A State-R$_k$ HOL packet would request to transmit with probability $q_0\mathcal{Q}(k)$ when sensing the channel idle, $k\in\{0,\cdots,K\}$. If its request is successful, it shifts to State T, stays in State T for $\tau_T^C$ time slots, and then shifts back to State R$_0$. Otherwise, it moves to State F$_k$, stays there for $\tau_{{F}_k}^C$ time slots, and then shifts to State R$_{\min\{k+1,K\}}$.

Similar to the Aloha case, for each node in the CSMA network, let $p_t^C$ denote the probability that its request of transmission is successful at time slot $t$ given that the channel is accessible and it requests at time slot $t$. The steady-state probability distribution of the embedded Markov chain $\mathbf{X^C}$ in Fig. \ref{fig3-2} can then be obtained as
\begin{equation}
    \begin{gathered}
        \pi_T^C = \tfrac{p^C}{2}, \pi_{R_k}^C=
        \begin{cases}
            (1-p^C)^k \pi_T^C & k\!\in\!\{0,\cdots,K\!-\!1\} \\
            \tfrac{(1-p^C)^K}{p^C}\pi_T^C & k\!=\!K,
        \end{cases} \\
        \pi_{F_k}^C=
        \begin{cases}
            (1-p^C)^{k+1}\pi_T^C & k\!\in\!\{0,\cdots,K\!-\!1\} \\
            \tfrac{(1-p^C)^{K+1}}{p^C}\pi_T^C & k\!=\!K,
        \end{cases}
    \end{gathered}
\label{eq3-1-2-1}
\end{equation}
where $p^C = \lim \limits_{t \to \infty} p_t^C$.

For CSMA, the mean holding time at State T can be written from (\ref{eq3-1-3}) as
\begin{equation}
    \tau_T^C =
    \begin{cases}
        \tfrac{L+\Delta_{S,P}}{\sigma_C} & {\text{for connection-free CSMA}} \\
        \tfrac{L+\Delta_{S,N}}{\sigma_C} & {\text{for connection-based CSMA}}.
    \end{cases}
\label{eq3-1-2-1.25}
\end{equation}
The holding time $Y_{{F}_k}^C$ at State F$_k$ is equal to the number of time slots of each failed transmission, $k\in\{0,\cdots,K\}$. Its mean $\tau_F^C$ can then be written as 
\begin{equation}
    \tau_{F}^C =
    \begin{cases}
        \tfrac{L+\Delta_{F,P}}{\sigma_C} & {\text{for connection-free CSMA}} \\
        \tfrac{\Delta_{F,N}}{\sigma_C} & {\text{for connection-based CSMA}}.
    \end{cases}
\label{eq3-1-2-1.5}
\end{equation}

When staying at State R$_k$, $k\in\{0,\cdots,K\}$, a HOL packet requests to transmit only if the channel is accessible. For each node, let $\tilde{\alpha}_t^C$ denote the probability that the channel is accessible (idle) at time slot $t$ given that it intends to request (i.e., its HOL packet is at State R$_k$, $k\in\{0,\cdots,K\}$) at time slot $t$. The holding time $Y_{{R}_k}^C$ at State R$_k$ can be modeled as a Geometric random variable with probability generating function:
\begin{equation}
    G_{Y_{{R}_k}^C}(z) = \tfrac{\tilde{\alpha}^C q_0\mathcal{Q}(k)z}{1-(1-\tilde{\alpha}^C q_0\mathcal{Q}(k))z},
\label{eq3-1-2-2}
\end{equation}
where $\tilde{\alpha}^C = \lim \limits_{t \to \infty} \tilde{\alpha}_t^C$. The mean holding time at State R$_k$ can then be obtained from (\ref{eq3-1-2-2}) as
\begin{equation}
    \tau_{{R}_k}^C = G_{Y_{{R}_k}^C}^\prime(1) = \tfrac{1}{\tilde{\alpha}^C q_0\mathcal{Q}(k)}.
\label{eq3-1-2-3}
\end{equation}

Finally, the limiting state probabilities of the Markov renewal process with CSMA $\{\tilde{\pi}_u^C\}_{u\in \mathcal{S}^C}$ can be obtained by combining (\ref{eq3-1-1}) with (\ref{eq3-1-2-1})-(\ref{eq3-1-2-1.5}), (\ref{eq3-1-2-3}), based on which the service rate of each node's queue $\mu^C$ can further be derived from (\ref{eq3-1-2}) as
\begin{equation}
    \mu^C \!=\! \tfrac{1}{\tau_T^C + \tau_F^C \tfrac{1-p^C}{p^C} + \tfrac{1}{p^C \tilde{\alpha}^C q_0}\Bigl(\!\tfrac{(1-p^C)^K}{\mathcal{Q}(K)} + \sum_{k=0}^{K-1}\tfrac{p^C(1-p^C)^k}{\mathcal{Q}(k)}\!\Bigr)}.
\label{eq3-1-2-5}
\end{equation}
\vspace{-0.7cm}
\subsection{Steady-State Probabilities $\tilde{\alpha}$ and $p$}\label{sec3-2}

It can be clearly seen from (\ref{eq3-1-1-5}) and (\ref{eq3-1-2-5}) that for both Aloha and CSMA, each node's steady-state performance is closely dependent on \romannumeral1) the steady-state probability that the channel is accessible given that the node intends to request, $\tilde{\alpha}$, and \romannumeral2) the steady-state probability that the request of transmission is successful given that the channel is accessible and the node requests, $p$, which will be derived in the following subsections.

\subsubsection{Steady-State Probability $\tilde{\alpha}$}\label{sec3-2-1}

Note that $\tilde{\alpha}$ is a conditional probability, which can be written as
\begin{equation}
    \tilde{\alpha} = \alpha\cdot\tfrac{\Pr\{{\rm The\,node\,intends\,to\,request}\,|\,{\rm Channel\,is\,accessible}\}}{\Pr\{{\rm The\,node\,intends\,to\,request}\}},
\label{eq3-2-1-1}
\end{equation}
where $\alpha$ denotes the unconditional steady-state probability that the channel is accessible to all the nodes. (\ref{eq3-2-1-1}) can be further written as
\begin{equation}
    \tilde{\alpha} =
    \begin{cases}
        \tfrac{\alpha}{1-\frac{\lambda}{\mu}+\frac{\lambda}{\mu}\sum_{u\in \mathcal{S}}\varphi_u} & \lambda<\mu \\
        \tfrac{\alpha}{\sum_{u\in \mathcal{S}}\varphi_u} & \lambda\geq\mu,
    \end{cases}
\label{eq3-2-1-2}
\end{equation}
where $\varphi_u$ denotes the probability that the node has a State-$u$ HOL packet and intends to request given that its queue is busy, $u\in\mathcal{S}$, which is given by
\begin{equation}
    \varphi_u^A =
    \begin{cases}
        \tilde{\pi}_{u}^A & u\in\{{\rm B}_0,\cdots,{\rm B}_K\} \\
        \tfrac{\tilde{\pi}_T^A}{\tau_T^A} & u\in\{{\rm T}\},
    \end{cases}
    \label{eq3-2-1-3}
\end{equation} 
and
\begin{equation}
    \varphi_u^C =
    \begin{cases}
        \tilde{\pi}_{u}^C & u\in\{{\rm R}_0,\cdots,{\rm R}_K\} \\
        0 & u\in\{{\rm T},{\rm F}_0,\cdots,{\rm F}_K\},
    \end{cases}
    \label{eq3-2-1-4}
\end{equation} 
for Aloha and CSMA, respectively. The unconditional probability of the channel being accessible $\alpha^A$ with Aloha and $\alpha^C$ with CSMA can be derived as
\begin{equation}
    \alpha^A \!=\! \tfrac{1}{1+np^A\bigl(1-(p^A)^{\frac{1}{n-1}}\bigr)\bigl(\tau_T^A-1\bigr)} \!\overset{n\gg1}{\approx}\! \tfrac{1}{1-(\tau_T^A-1)p^A\ln p^A},
\label{eq3-2-1-5}
\end{equation}
and
\begin{equation}
    \begin{split}
        \alpha^C & = \tfrac{1}{1+\tau_F^C\bigl(1-(p^C)^{\frac{n}{n-1}}\bigr)+np^C(\tau_T^C-\tau_F^C)\bigl(1-(p^C)^{\frac{1}{n-1}}\bigr)} \\
        & \overset{n\gg1}{\approx} \tfrac{1}{1 + \tau_F^C(1-p^C) -(\tau_T^C-\tau_F^C)p^C\ln p^C},
    \end{split}
\label{eq3-2-1-6}
\end{equation}
respectively, where the approximation is based on $n-1\approx n$ and $n(1-x^{\frac{1}{n}})\approx-\ln x$ for $0<x<1$ if $n\gg 1$. The detailed derivation of (\ref{eq3-2-1-2})-(\ref{eq3-2-1-6}) is shown in Appendix \ref{sec_ap1}.

Finally, by combining (\ref{eq3-2-1-2})-(\ref{eq3-2-1-3}), (\ref{eq3-2-1-5}) with (\ref{eq3-1-1}), (\ref{eq3-1-1-1}), (\ref{eq3-1-1-3})-(\ref{eq3-1-1-5}), we have
\vspace{0.2cm}
\begin{equation}
    \tilde{\alpha}^A =
    \begin{cases}
        \tfrac{1}{1-\lambda(\tau_T^A-1)}\cdot\tfrac{1}{1-(\tau_T^A-1)p^A\ln p^A} & \lambda<\mu^A \\
        \tfrac{1}{1-(\tau_T^A-1)p^A\ln p^A-(\tau_T^A-1)\cdot\tfrac{p^A q_0}{f(p^A)}} & \lambda\geq\mu^A,
    \end{cases}
\label{eq3-2-1-7}
\end{equation}
for Aloha, and
\vspace{0.2cm}
\begin{equation}
    \tilde{\alpha}^C =
    \begin{cases}
        \tfrac{1}{1-\lambda\Big(\tau_T^C+\tau_F^C\frac{1-p^C}{p^C}\Big)}\cdot\tfrac{1}{1 + \tau_F^C(1-p^C) -(\tau_T^C-\tau_F^C)p^C\ln p^C} \\
        \hspace{5cm} \lambda<\mu^C \\
        \tfrac{1}{1 + \tau_F^C(1-p^C) -(\tau_T^C-\tau_F^C)p^C\ln p^C-\big(\tau_T^C +\tau_F^C\frac{1-p^C}{p^C}\big)\cdot\tfrac{p^C q_0}{f(p^C)}} \\
        \hspace{5cm} \lambda\geq\mu^C,
    \end{cases}
\label{eq3-2-1-8}
\end{equation}
for CSMA by combining (\ref{eq3-2-1-2}), (\ref{eq3-2-1-4}), (\ref{eq3-2-1-6}) with (\ref{eq3-1-1}), (\ref{eq3-1-2-1}), (\ref{eq3-1-2-3})-(\ref{eq3-1-2-5}), where $f(p) = \tfrac{(1-p)^K}{\mathcal{Q}(K)} + \sum_{k=0}^{K-1}\tfrac{p(1-p)^k}{\mathcal{Q}(k)}$.
\vspace{0.2cm}
\subsubsection{Steady-State Probability $p$}\label{sec3-2-2}

With the collision model, a transmission is successful only when there are no concurrent transmissions. $p$ can then be written as
\vspace{0.2cm}
\begin{equation}
    \begin{split}
        p = &\Pr\{{\rm All\,the\,other\,}n-1\,{\rm nodes\,do\,not\,request\,}| \\
        & {\rm Channel\,is\,accessible\,and\,the\,node\,requests}\} \\
        = & (1-\omega)^{n-1},
    \end{split}
\label{eq3-2-2-1}
\end{equation}
where $\omega = \Pr\{{\rm The\,node\,has\,a\,busy\,queue\,and\,makes\,a\,}$ ${\rm request}\,|\,{\rm Channel\,is\,accessible}\}$ can be obtained as
\vspace{0.2cm}
\begin{equation}
    \omega =
    \begin{cases}
        \tfrac{\lambda}{p\alpha} & \lambda<\mu \\
        \tfrac{\mu}{p\alpha} & \lambda\geq\mu.
    \end{cases}
    \label{eq3-2-2-2}
\end{equation}
The detailed derivation of (\ref{eq3-2-2-2}) is shown in Appendix \ref{sec_ap2}. It is clear from (\ref{eq3-2-2-1})-(\ref{eq3-2-2-2}) that the steady-state probability of successful transmission of HOL packets $p$ crucially depends on whether the node's queue is saturated or not.

Specifically, in the unsaturated condition with $\lambda<\mu$, (\ref{eq3-2-2-1}) can be written as
\begin{equation}
    p = (1-\tfrac{\lambda}{p\alpha})^{n-1}.
\label{eq3-2-2-3}
\end{equation}
With Aloha, the fixed-point equation of $p$ can be established by combining (\ref{eq3-2-2-3}) with (\ref{eq3-2-1-5}) as 
\begin{equation}
    p=\Bigl(1-\tfrac{\lambda}{p (1-\hat{\lambda}(\tau_T^A-1))}\Bigr)^{n-1} \overset{n\gg 1}{\approx} e^{-\frac{\hat{\lambda}}{p(1-\hat{\lambda}(\tau_T^A-1))}},
\label{eq3-2-2-4}
\end{equation}
where the approximation is based on $n-1\approx n$ and $(1-x)^n\approx e^{-nx}$ for $0<x<1$ if $n\gg 1$. The fixed-point equation (\ref{eq3-2-2-4}) has two positive real roots:
\begin{equation}
    p_L^A = e^{\mathbb{W}_0\Bigl(-\frac{\hat{\lambda}}{1-\hat{\lambda}(\tau_T^A-1)}\Bigr)},\,p_S^A = e^{\mathbb{W}_{-1}\Bigl(-\frac{\hat{\lambda}}{1-\hat{\lambda}(\tau_T^A-1)}\Bigr)},
\label{eq3-2-2-5}
\end{equation}
if and only if
\begin{equation}
    \tfrac{\hat{\lambda}}{1-\hat{\lambda}(\tau_T^A-1)} < e^{-1},
\label{eq3-2-2-5.5}
\end{equation}
where $\mathbb{W}_0(\cdot)$ and $\mathbb{W}_{-1}(\cdot)$ are two branches of the Lambert W function \cite{corless1996lambert}. With CSMA, the fixed-point equation of $p$ can be established by combining (\ref{eq3-2-2-3}) with (\ref{eq3-2-1-6}) as
\begin{equation}
    \begin{split}
        p & = \Bigl(1\!+\!\tfrac{\lambda \tau_F^C}{1-\hat{\lambda}\tau_T^C+\lambda\tau_F^C(n-1)}\!-\!\tfrac{\lambda(\tau_F^C+1)}{p(1-\hat{\lambda}\tau_T^C+\lambda\tau_F^C(n-1))}\Bigr)^{n-1} \\
        \! & \overset{n\gg 1}{\approx} e^{\frac{\hat{\lambda}\tau_F^C}{1-\hat{\lambda}(\tau_T^C-\tau_F^C)}-\frac{\hat{\lambda}(\tau_F^C+1)}{p(1-\hat{\lambda}(\tau_T^C-\tau_F^C))}}.
    \end{split}
\label{eq3-2-2-6}
\end{equation}
The fixed-point equation (\ref{eq3-2-2-6}) also has two positive real roots:
\begin{equation}
    \begin{split}
        & p_L^C = e^{\mathbb{W}_0\Bigl(-\frac{\hat{\lambda} (\tau_F^C+1)}{1-\hat{\lambda}(\tau_T^C-\tau_F^C)}\cdot e^{-\frac{\hat{\lambda} \tau_F^C}{1-\hat{\lambda}(\tau_T^C-\tau_F^C)}}\Bigr)+\frac{\hat{\lambda} \tau_F^C}{1-\hat{\lambda}(\tau_T^C-\tau_F^C)}}, \\
        & p_S^C = e^{\mathbb{W}_{-1}\Bigl(-\frac{\hat{\lambda} (\tau_F^C+1)}{1-\hat{\lambda}(\tau_T^C-\tau_F^C)}\cdot e^{-\frac{\hat{\lambda} \tau_F^C}{1-\hat{\lambda}(\tau_T^C-\tau_F^C)}}\Bigr)+\frac{\hat{\lambda} \tau_F^C}{1-\hat{\lambda}(\tau_T^C-\tau_F^C)}},
    \end{split}
\label{eq3-2-2-7}
\end{equation}
if and only if
\begin{equation}
    \tfrac{\hat{\lambda} (\tau_F^C+1)}{1-\hat{\lambda}(\tau_T^C-\tau_F^C)}\cdot e^{-\frac{\hat{\lambda} \tau_F^C}{1-\hat{\lambda}(\tau_T^C-\tau_F^C)}} < e^{-1}.
\label{eq3-2-2-7.5}
\end{equation}
Note that for both the fixed-point equations (\ref{eq3-2-2-4}) and (\ref{eq3-2-2-6}), the root $p_L$ is an attracting fixed point, while $p_S$ is a repelling fixed point. Therefore, only the larger root $p_L$ is the steady-state point in the unsaturated condition. It can be seen from (\ref{eq3-2-2-5}) and (\ref{eq3-2-2-7}) that for both Aloha and CSMA, $p_L$ is closely determined by the aggregate packet input rate $\hat{\lambda}$. Note that with $\tau_T^A=1$ for connection-free Aloha and $\tau_T^A=\tfrac{L+\Delta_{S,N}}{\Delta_{F,N}}$ for connection-based Aloha, (\ref{eq3-2-2-5}) is consistent with Eq. (80) in \cite{9765641} and Eq. (18)-(19) in \cite{10154598}, respectively;\footnote{Note that by using the notations in this paper, we can rewrite $n\lambda_r$, $M$ and $\delta$ in Eq. (18)-(19) of \cite{10154598} as $\hat{\lambda}$, $\tfrac{L}{\Delta_{F,N}}$ and $\tfrac{\Delta_{S,N}}{\Delta_{F,N}}$, respectively, with which Eq. (18) and Eq. (19) in \cite{10154598} are same as (\ref{eq3-2-2-5}) in this paper.} (\ref{eq3-2-2-7}) for CSMA is consistent with Eq. (34)-(35) in \cite{6525472}.

On the other hand, in the saturated condition with $\lambda\geq\mu$, the fixed-point equation of $p$ can be established based on (\ref{eq3-2-2-1})-(\ref{eq3-2-2-2}) as
\begin{equation}
    p = (1-\tfrac{\mu}{p\alpha})^{n-1}.
\label{eq3-2-2-8}
\end{equation}
Note that for Aloha and CSMA, the service rate of each node's queue in the saturated condition can be written from (\ref{eq3-1-1-5}), (\ref{eq3-2-1-5}), (\ref{eq3-2-1-7}) and (\ref{eq3-1-2-5}), (\ref{eq3-2-1-6}), (\ref{eq3-2-1-8}) as
\begin{equation}
    \mu^A=\tfrac{p^A\alpha^A q_0}{f(p^A)}\,\,\,\text{and}\,\,\,\mu^C=\tfrac{p^C\alpha^C q_0}{f(p^C)},
\label{eq3-2-2-8.5}
\end{equation}
respectively, where $f(p) = \tfrac{(1-p)^K}{\mathcal{Q}(K)} + \sum_{k=0}^{K-1}\tfrac{p(1-p)^k}{\mathcal{Q}(k)}$. By combining (\ref{eq3-2-2-8}) with (\ref{eq3-2-2-8.5}), we have
\begin{equation}
    \begin{split}
        p \!=\! \biggl(\!1\!-\!\tfrac{q_0}{\frac{(1-p)^K}{\mathcal{Q}(K)} + \sum_{k=0}^{K-1}\frac{p(1-p)^k}{\mathcal{Q}(k)}}\!\biggr)^{n-1} \!\overset{n\gg 1}{\approx}\! e^{\frac{-n q_0}{\frac{(1-p)^K}{\mathcal{Q}(K)} + \sum_{k=0}^{K-1}\frac{p(1-p)^k}{\mathcal{Q}(k)}}}\!,
    \end{split}
    \label{eq3-2-2-9}
\end{equation}
for both Aloha and CSMA, where the approximation is based on $n-1\approx n$ and $(1-x)^n\approx e^{-nx}$ for $0<x<1$ if $n\gg 1$. It has been shown in \cite{9765641} that the fixed-point equation (\ref{eq3-2-2-9}) has a single root in $(0,1]$ as long as the backoff function $\mathcal{Q}(k)$ is a monotonic non-increasing function of $k$. This root, denoted as $p_A$, is the steady-state point in the saturated condition and determined by the number of nodes $n$, the initial transmission probability $q_0$ and backoff function $\mathcal{Q}(k)$. Note that (\ref{eq3-2-2-9}) is consistent with Eq. (97) in \cite{9765641} and Eq. (51) in \cite{6525472}.

\subsection{Unsaturated Region of Initial Transmission Probability}\label{sec3-3}

We can see from the analysis above that each node's steady-state performance is crucially determined by whether its queue is saturated or not, which further depends on the initial transmission probability $q_0$.

Specifically, each node's queue would be saturated if $\lambda\geq \mu$, where $\mu=\tfrac{p_A\alpha q_0}{f(p_A)}$ in the saturated condition according to (\ref{eq3-2-2-8.5}). For each node's queue to stay unsaturated, the initial transmission probability $q_0$ should satisfy $\lambda<\tfrac{p_A\alpha q_0}{f(p_A)}$. According to (\ref{eq3-2-2-9}), $\lambda<\tfrac{p_A\alpha q_0}{f(p_A)}$ can be further written as $p_A<(1-\tfrac{\lambda}{p_A\alpha})^{n-1}$, which requires $p_S<p_A<p_L$, where $p_L$ and $p_S$ are two positive real roots of the fixed-point equation (\ref{eq3-2-2-3}). By noting that $p_A$ is a function of $q_0$ given by (\ref{eq3-2-2-9}), the unsaturated region of initial transmission probability $S_{U^n}(q_0, \hat{\lambda})$ can be obtained as
\begin{equation}
    \begin{split}
        & S_{U^n}(q_0, \hat{\lambda}) \!=\! \biggl\{q_0: -\tfrac{\ln p_L}{n}\biggl(\!\tfrac{(1-p_L)^K}{\mathcal{Q}(K)} \!+\! \sum_{k=0}^{K-1}\tfrac{p_L(1-p_L)^k}{\mathcal{Q}(k)}\!\biggr) \\
        & < q_0 < -\tfrac{\ln p_S}{n}\biggl(\!\tfrac{(1-p_S)^K}{\mathcal{Q}(K)} \!+\! \sum_{k=0}^{K-1}\tfrac{p_S(1-p_S)^k}{\mathcal{Q}(k)}\!\biggr)\biggr\},
    \end{split}
    \label{eq3-3-1}
\end{equation}
where $p_L$ and $p_S$ are given by (\ref{eq3-2-2-5}) for Aloha, and (\ref{eq3-2-2-7}) for CSMA.

Note from (\ref{eq3-3-1}) that $S_{U^n}(q_0, \hat{\lambda})$ is nonempty only when $p_L$ and $p_S$ exist, which requires that the aggregate input rate of data packets $\hat{\lambda}$ satisfies (\ref{eq3-2-2-5.5}) for Aloha and (\ref{eq3-2-2-7.5}) for CSMA. (\ref{eq3-2-2-5.5}) and (\ref{eq3-2-2-7.5}) can be rewritten as $\hat{\lambda}<\hat{\lambda}_{\max}^A$ and $\hat{\lambda}<\hat{\lambda}_{\max}^C$, respectively, where
\begin{equation}
    \hat{\lambda}_{\max}^A = \tfrac{1}{\tau_T^A-1+e}
    \label{eq3-2-2}
\end{equation}
and
\begin{equation}
    \hat{\lambda}_{\max}^C = \tfrac{-\mathbb{W}_0\bigl(-\tfrac{\tau_F^C}{e(\tau_F^C+1)}\bigr)}{\tau_F^C-(\tau_T^C-\tau_F^C)\mathbb{W}_0\bigl(-\tfrac{\tau_F^C}{e(\tau_F^C+1)}\bigr)}
    \label{eq3-2-3}
\end{equation}
are indeed the maximum data throughput with Aloha and CSMA, respectively.

\section{Minimum Mean Queueing Delay of Data Packets}\label{sec4}

Based on the HOL-packet model in Section \ref{sec3}, the mean queueing delay of data packets $\overline{T}$ will be characterized and optimized in this section. Specifically, with Bernoulli arrival process of data packets, $\overline{T}$ can be written as
\begin{equation}
    \overline{T} = \tfrac{\lambda \overline{D}^2-\lambda \overline{D}}{2(1-\lambda \overline{D})}+\overline{D},
\label{eq4-1-1}
\end{equation}
for $\lambda\overline{D}<1$, where $\overline{D}$ and $\overline{D}^2$ are the first and second moments of the service time of each node's queue, respectively. Note that $\overline{D}=\tfrac{1}{\mu}$, and thus $\lambda\overline{D}<1$ if and only if each node's queue is unsaturated. For each node's queue to stay unsaturated, the initial transmission probability $q_0$ should be selected from the unsaturated region $S_{U^n}(q_0,\hat{\lambda})$. With $q_0\in S_{U^n}(q_0,\hat{\lambda})$, the mean queueing delay of data packets $\overline{T}$ can further be minimized by optimally tuning $q_0$.

It can be seen that to characterize and optimize the mean queueing delay of data packets, the key lies in the derivation of the first moment $\overline{D}$ and second moment $\overline{D}^2$ of service time of each node's queue. In the following subsection, $\overline{D}$ and $\overline{D}^2$ will be obtained by characterizing the service time distribution of each node's queue based on the Markov renewal process of HOL packets established in Section \ref{sec3}.

\subsection{First and Second Moments of Service Time}\label{sec4-1}

\begin{figure*}[!t]
    \normalsize
    \setcounter{mytempeqncnt}{\value{equation}}
    \setcounter{equation}{51}
    \begin{equation}
        \begin{split}
            G_{D^A}^{\prime\prime}(1) = & \sum_{i=0}^{K-1} \tfrac{2(1-p^A)^i}{\tilde{\alpha}^A q_0\mathcal{Q}(i)}\bigl(\tfrac{1}{\tilde{\alpha}^A q_0\mathcal{Q}(i)}+\tau_T^A-2\bigr) + \sum_{i=0}^{K-1} \tfrac{2}{\tilde{\alpha}^A q_0\mathcal{Q}(i)}\sum_{j=i+1}^{K-1}\tfrac{(1-p^A)^j}{\tilde{\alpha}^A q_0\mathcal{Q}(j)} + \sum_{i=0}^{K-1}\tfrac{2(1-p^A)^K}{p^A(\tilde{\alpha}^A q_0)^2\mathcal{Q}(K)\mathcal{Q}(i)} \\
            & + \tfrac{2(1-p^A)^{K}}{p^A\tilde{\alpha}^A q_0\mathcal{Q}(K)}\bigl(\tfrac{1}{p^A\tilde{\alpha}^A q_0\mathcal{Q}(K)}+\tau_T^A-2\bigr) + (\tau_T^A-1)(\tau_T^A-2).
        \end{split}
    \label{eq4-1-1-5}
    \end{equation}
    \setcounter{equation}{53}
    \begin{equation}
        \begin{split}
            G_{D^C}^{\prime\prime}(1) = & \sum_{i=0}^{K-1}2(\tau_F^C+\tfrac{1}{\tilde{\alpha}^Cq_0\mathcal{Q}(i)})\cdot\biggl(\sum_{j=i+1}^{K-1}\tfrac{(1-p^C)^j}{\tilde{\alpha}^Cq_0\mathcal{Q}(j)} + (\tau_T^C +\tau_F^C\tfrac{1-p^C}{p^C})(1-p^C)^{i+1} + \tfrac{(1-p^C)^K}{p^C\tilde{\alpha}^Cq_0\mathcal{Q}(K)}\biggr) \\
            & + \sum_{i=0}^{K-1} \tfrac{2p^C(1-p^C)^i}{\tilde{\alpha}^Cq_0\mathcal{Q}(i)}\bigl(\tau_T^C +\tau_F^C\tfrac{1-p^C}{p^C}+\tfrac{1}{p^C\tilde{\alpha}^Cq_0\mathcal{Q}(i)}-\tfrac{1}{p^C}\bigr) + \tau_T^C(\tau_T^C-1) + \tau_F^C(\tau_F^C-1)\tfrac{1-p^C}{p^C} \\
            & + 2(1-p^C)^K \Bigl(\tfrac{1-p^C}{p^C}\tau_F^C(\tau_T^C + \tau_F^C\tfrac{1-p^C}{p^C})+\tfrac{1}{p^C\tilde{\alpha}^Cq_0\mathcal{Q}(K)}(\tau_T^C+\tfrac{2(1-p^C)}{p^C}\tau_F^C-1)+\tfrac{1}{(p^C\tilde{\alpha}^Cq_0\mathcal{Q}(K))^2}\Bigr) .
        \end{split}
    \label{eq4-1-1-7}
    \end{equation}
    \setcounter{equation}{\value{mytempeqncnt}}
    \hrulefill
    \vspace{-0.2cm}
\end{figure*}

The first moment $\overline{D}$ and second moment $\overline{D}^2$ of service time of each node's queue can be written as
\begin{equation}
    \overline{D} = G_{D}^\prime(1)\;\;\;{\rm{and}}\;\;\;\overline{D}^2 = G_{D}^{\prime\prime}(1)+G_{D}^\prime(1),
\label{eq4-1-1-1}
\end{equation}
respectively, where $G_D(z)$ denotes the probability generating function of service time of each queue. In general, given the embedded Markov chain of each HOL packet's state transition process and holding time distribution at each state, the probability generating function of service time $G_D(z)$ can be obtained as
\begin{equation}
    G_D(z) = \sum_{u\in\mathcal{S}} P_{T,u} G_{D_u}(z),
\label{eq4-1-1-2}
\end{equation}
where $P_{T,u}$ denotes the transition probability from State T to State $u$ in the embedded Markov chain $\mathbf{X}$, and $G_{D_u}(z)$ denotes the probability generating function of $D_u$, i.e., the number of time slots from a HOL packet shifting to State $u$ till the service completion. $G_{D_u}(z)$ can further be written as
\begin{equation}
    G_{D_u}(z) = 
    \begin{cases}
        z^{\tau_T} & u\in\{{\rm{T}}\} \\
        \sum_{v\in\mathcal{S}} P_{u,v}G_{Y_u}(z)G_{D_v}(z) & u\in\mathcal{S}\backslash\{{\rm{T}}\},
    \end{cases}
\label{eq4-1-1-3}
\end{equation}
where $P_{u,v}$ is the transition probability from State $u$ to State $v$ in the embedded Markov chain $\mathbf{X}$, and $G_{Y_u}(z)$ is the probability generating function of the holding time at State $u$. The derivation of (\ref{eq4-1-1-2})-(\ref{eq4-1-1-3}) is presented in Appendix \ref{sec_ap3}.

For Aloha and CSMA, the embedded Markov chain of each HOL packet's state transition process is presented in Fig. \ref{fig3-1} and Fig. \ref{fig3-2}, respectively, and the holding time distribution at each state is given by (\ref{eq3-1-1-2}) and (\ref{eq3-1-2-2}), respectively. By combining with (\ref{eq4-1-1-2})-(\ref{eq4-1-1-3}), the probability generating function of service time of each node's queue $G_{D^A}(z)$ with Aloha and $G_{D^C}(z)$ with CSMA can be obtained. Based on that, $G_{D^A}^{\prime}(1)$ and $G_{D^A}^{\prime\prime}(1)$ for Aloha can be derived as
\vspace{-0.1cm}
\begin{equation}
    G_{D^A}^{\prime}(1) = \tau_T^A - 1 + \tfrac{(1-p^A)^K}{p^A\tilde{\alpha}^A q_0\mathcal{Q}(K)} + \sum_{i=0}^{K-1} \tfrac{(1-p^A)^i}{\tilde{\alpha}^A q_0\mathcal{Q}(i)},
\label{eq4-1-1-4}
\end{equation}
and (\ref{eq4-1-1-5}), respectively, where $\tilde{\alpha}^A$ is given by (\ref{eq3-2-1-7}), and $p^A$ is given by (\ref{eq3-2-2-5}) and (\ref{eq3-2-2-9}) in the unsaturated and saturated conditions, respectively. (\ref{eq4-1-1-5}) is shown at the top of this page. For CSMA, $G_{D^C}^{\prime}(1)$ and $G_{D^C}^{\prime\prime}(1)$ can be derived as
\setcounter{equation}{52}
\begin{equation}
    G_{D^C}^{\prime}(1) \!=\! \tau_T^C \!+\! \tau_F^C\tfrac{1-p^C}{p^C} \!+\! \tfrac{(1-p^C)^K}{p^C\tilde{\alpha}^Cq_0\mathcal{Q}(K)} \!+\! \sum_{i=0}^{K-1} \tfrac{(1-p^C)^i}{\tilde{\alpha}^Cq_0\mathcal{Q}(i)},
\label{eq4-1-1-6}
\end{equation}
and (\ref{eq4-1-1-7}), respectively, where $\tilde{\alpha}^C$ is given by (\ref{eq3-2-1-8}), and $p^C$ is given by (\ref{eq3-2-2-7}) and (\ref{eq3-2-2-9}) in the unsaturated and saturated conditions, respectively. (\ref{eq4-1-1-7}) is shown at the top of this page. The first moment $\overline{D}$ and second moment $\overline{D}^2$ of service time of each node's queue can then be derived by combining (\ref{eq4-1-1-1}) with (\ref{eq4-1-1-4})-(\ref{eq4-1-1-5}) for Aloha, and with (\ref{eq4-1-1-6})-(\ref{eq4-1-1-7}) for CSMA.

\subsection{Effect of Backoff Function on Minimum Mean Queueing Delay}\label{sec4-2}

Based on the analysis above, the mean queueing delay of data packets $\overline{T}$ can be derived by combining (\ref{eq4-1-1})-(\ref{eq4-1-1-1}) with (\ref{eq3-2-1-7}), (\ref{eq3-2-2-5}), (\ref{eq4-1-1-4})-(\ref{eq4-1-1-5}) for Aloha, and with (\ref{eq3-2-1-8}), (\ref{eq3-2-2-7}), (\ref{eq4-1-1-6})-(\ref{eq4-1-1-7}) for CSMA. As both $\overline{T}^A$ with Aloha and $\overline{T}^C$ with CSMA are monotonic decreasing functions of the initial transmission probability $q_0$ for $q_0\in S_{U^n}(q_0, \hat{\lambda})$, according to (\ref{eq3-3-1}), the optimal initial transmission probability $q_0^{\ast}$ for minimizing the mean queueing delay can be obtained as
\setcounter{equation}{54}
\vspace{0.1cm}
\begin{equation}
    q_0^{\ast} = -\tfrac{\ln p_S}{n}\biggl(\tfrac{(1-p_S)^K}{\mathcal{Q}(K)} + \sum_{k=0}^{K-1}\tfrac{p_S(1-p_S)^k}{\mathcal{Q}(k)}\biggr) - \epsilon,
\label{eq4-2-1}
\end{equation}
where $p_S$ is given by (\ref{eq3-2-2-5}) for Aloha and (\ref{eq3-2-2-7}) for CSMA, and $\epsilon$ is a positive small value.

We can see from (\ref{eq4-2-1}) that the optimal delay performance is closely dependent on the backoff function $\mathcal{Q}(k)$, $k\in\{0,\cdots,K\}$. For example, with Constant Backoff (CB), the backoff function is $\mathcal{Q}(k)=1$, $k\in\{0,\cdots,K\}$. Another widely adopted backoff scheme is Binary Exponential Backoff (BEB), with which the backoff function is given by $\mathcal{Q}(k)=2^{-k}$, $k\in\{0,\cdots,K\}$. Given the backoff function $\mathcal{Q}(k)$, the mean queueing delay of data packets $\overline{T}$ can be derived by combining $\mathcal{Q}(k)$ and (\ref{eq4-1-1})-(\ref{eq4-1-1-1}) with (\ref{eq3-2-1-7}), (\ref{eq3-2-2-5}), (\ref{eq4-1-1-4})-(\ref{eq4-1-1-5}) for Aloha, and with (\ref{eq3-2-1-8}), (\ref{eq3-2-2-7}), (\ref{eq4-1-1-6})-(\ref{eq4-1-1-7}) for CSMA.

With CB, the unsaturated region of initial transmission probability can be obtained by substituting $\mathcal{Q}(k)=1$ into (\ref{eq3-3-1}) as
\vspace{0.1cm}
\begin{equation}
    S_{U^n}^{CB}(q_0, \hat{\lambda}) = \Bigl\{q_0:-\tfrac{1}{n}\ln p_L <q_0<-\tfrac{1}{n}\ln p_S\Bigr\}.
\label{eq4-2-1-2}
\end{equation}
With the initial transmission probability $q_0\in S_{U^n}^{CB}(q_0, \hat{\lambda})$, the mean queueing delay of data packets $\overline{T}^{CB}$ with CB is minimized when $q_0$ is set to:
\vspace{0.1cm}
\begin{equation}
    q_0^{\ast,CB} = -\tfrac{1}{n}\ln p_S-\epsilon,
\label{eq4-2-1-3}
\end{equation}
where $\epsilon$ is a positive small value. The minimum mean queueing delay of data packets $\overline{T}_{\min}^{CB}$ with CB can then be obtained by substituting (\ref{eq4-2-1-3}) into $\overline{T}^{CB}$.

\begin{figure*}[t]
    \centering
    \subfloat[]{
        \includegraphics[width=0.39\textwidth, height=0.23\textwidth]{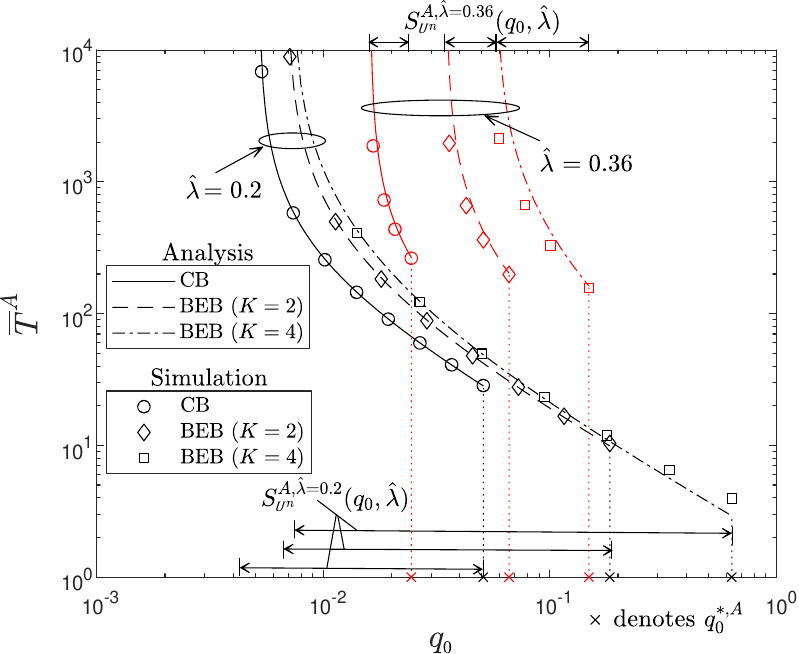}
        \label{fig4-1-1}
    }\hspace{1.2cm}
    \subfloat[]{
        \includegraphics[width=0.39\textwidth, height=0.23\textwidth]{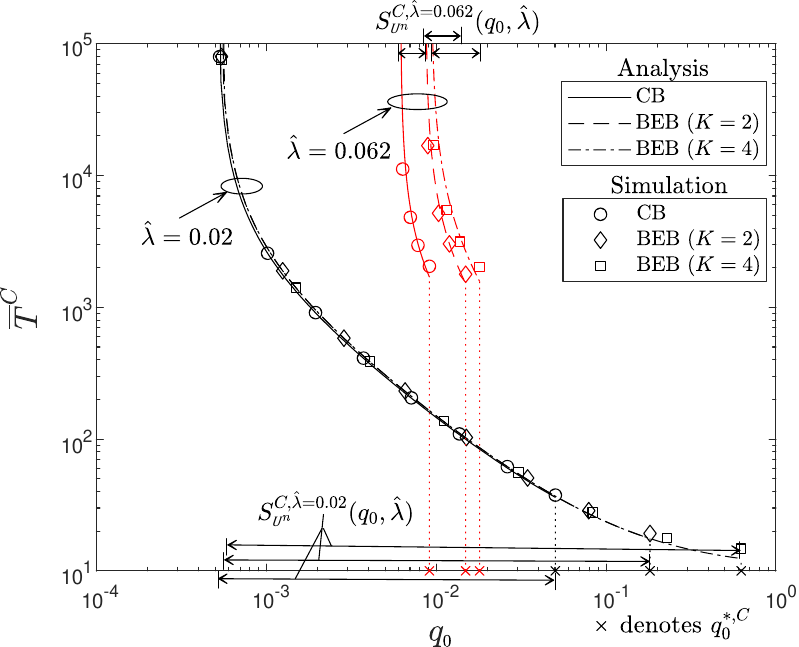}
        \label{fig4-1-2}
    }
    \caption{Analytical and simulated mean queueing delay of data packets $\overline{T}$ (in the unit of time slots) versus initial transmission probability $q_0$. $n=50$. (a) Aloha. $\hat{\lambda}\in\{0.2,0.36\}$. $\tau_T^A=1$. (b) CSMA. $\hat{\lambda}\in\{0.02,0.062\}$. $\tau_T^C=\tau_F^C=10$.}
    \label{fig4-1}
    \vspace{-0.3cm}
\end{figure*}

For BEB, on the other hand, the unsaturated region of initial transmission probability $S_{U^n}^{BEB}(q_0, \hat{\lambda})$ can be obtained by substituting $\mathcal{Q}(k)=2^{-k}$ into (\ref{eq3-3-1}) as
\begin{equation}
    \begin{split}
        & S_{U^n}^{BEB}(q_0, \hat{\lambda}) = \Bigl\{q_0:-\tfrac{\ln p_L}{n}\bigl(\tfrac{2^K(1-p_L)^{K+1}}{1-2p_L}-\tfrac{p_L}{1-2p_L}\bigr) \\
            & <q_0<-\tfrac{\ln p_S}{n}\bigl(\tfrac{2^K(1-p_S)^{K+1}}{1-2p_S}-\tfrac{p_S}{1-2p_S}\bigr)\Bigr\}.
    \end{split}
    \label{eq4-2-2-1}
\end{equation}
With the initial transmission probability $q_0\in S_{U^n}^{BEB}(q_0, \hat{\lambda})$, the mean queueing delay of data packets $\overline{T}^{BEB}$ with BEB is minimized when $q_0$ is set to:
\begin{equation}
    q_0^{\ast,BEB} = -\tfrac{\ln p_S}{n}\bigl(\tfrac{2^K(1-p_S)^{K+1}}{1-2p_S}-\tfrac{p_S}{1-2p_S}\bigr)-\epsilon,
    \label{eq4-2-2-2}
\end{equation}
where $\epsilon$ is a positive small value. The minimum mean queueing delay of data packets $\overline{T}_{\min}^{BEB}$ with BEB can then be obtained by substituting (\ref{eq4-2-2-2}) into $\overline{T}^{BEB}$.

Fig. \ref{fig4-1} illustrates how the mean queueing delay of data packets $\overline{T}$ varies with the initial transmission probability $q_0$ under different settings of backoff scheme and aggregate input rate of data packets $\hat{\lambda}$. It can be seen that $\overline{T}$ is finite only when $q_0$ is chosen from the unsaturated region $S_{U^n}(q_0, \hat{\lambda})$. For $q_0\in S_{U^n}(q_0, \hat{\lambda})$, $\overline{T}$ monotonically decreases with $q_0$, and is minimized when $q_0$ is set to $q_0^{\ast}$. With a larger aggregate packet input rate $\hat{\lambda}$, the unsaturated region of initial transmission probability $S_{U^n}(q_0, \hat{\lambda})$ becomes smaller, and a lower initial transmission probability $q_0$ should be selected for minimizing the mean queueing delay. To verify the analysis, simulation results are also presented in Fig. \ref{fig4-1}. The simulation setting is identical to the system model described in Section \ref{sec2}. Each simulation is conducted for $10^8$ time slots. In simulations, the mean queueing delay of data packets $\overline{T}$ is obtained by calculating the ratio of mean queue length (i.e., average number of data packets in each node's queue per time slot) to packet input rate of each node. It can be seen from Fig. \ref{fig4-1} that the analytical results well agree with the simulation results.

It can also be observed from Fig. \ref{fig4-1} that for the given initial transmission probability $q_0$, the mean queueing delay of data packets $\overline{T}^{BEB}$ with BEB is higher than $\overline{T}^{CB}$ with CB. However, when $q_0$ is optimally selected for minimizing the mean queueing delay, owing to a larger optimal initial transmission probability, the network with BEB can indeed achieve a lower minimum mean queueing delay than that with CB. Furthermore, we can see from Fig. \ref{fig4-1} that the mean queueing delay of data packets $\overline{T}$ also crucially depends on which type of random access is adopted, Aloha or CSMA. In the next section, the optimal delay performance of Aloha and CSMA will be compared. Note that $\overline{T}$ is in the unit of time slots. However, the time slot length $\sigma_A$ with Aloha and $\sigma_C$ with CSMA, in the unit of ms, are usually unequal. Therefore, in the following comparative study, we will focus on the minimum mean queueing delay of data packets in the unit of ms $\tilde{T}$.

\section{Sensing-Free Versus Sensing-Based Random Access}\label{sec5}

Whether sensing can improve the optimal delay performance of random access is crucially determined by the sensing time $\sigma_C$. To be more specific, CSMA outperforms Aloha in terms of the minimum mean queueing delay of data packets $\tilde{T}_{\min}$ in the unit of ms only when $\sigma_C$ is below the upper-bound $\tilde{\sigma}_C^{\ast}$, which will be characterized in the following subsection.

\subsection{Delay-Optimal Sensing Bound}\label{sec5-1}

\begin{figure*}[t]
    \centering
    \subfloat[]{
        \includegraphics[width=0.39\textwidth, height=0.22\textwidth]{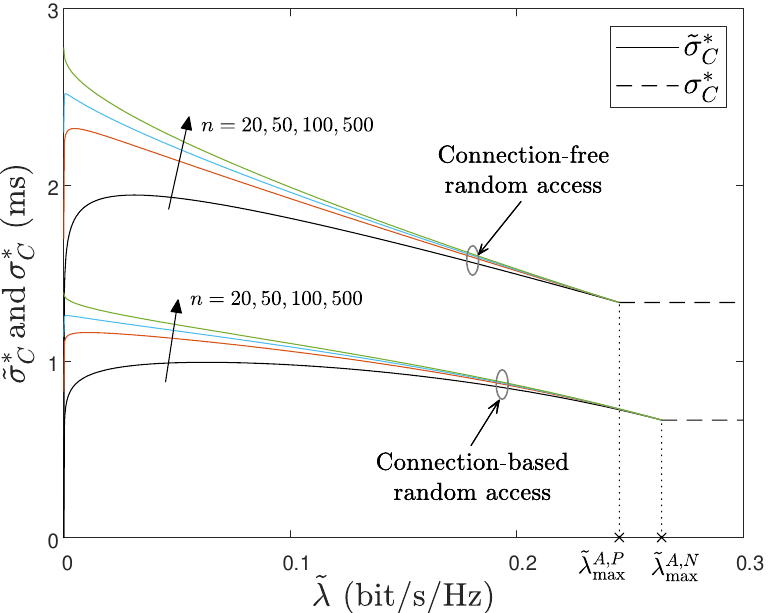}
        \label{fig5-1-1}
    }\hspace{1.2cm}
    \subfloat[]{
        \includegraphics[width=0.39\textwidth, height=0.22\textwidth]{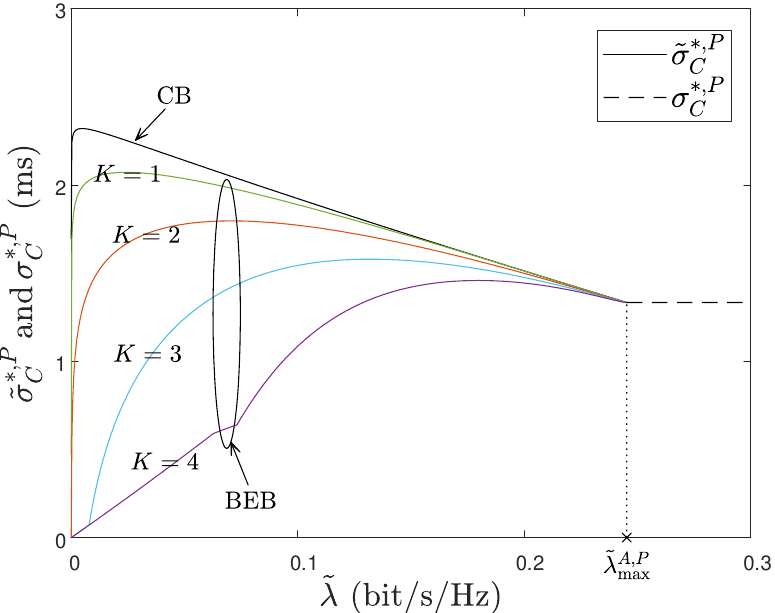}
        \label{fig5-1-2}
    }
    \caption{Delay-optimal sensing bound $\tilde{\sigma}_C^{\ast}$ and throughput-optimal sensing bound $\sigma_C^{\ast}$ versus aggregate bit input rate $\tilde{\lambda}$. $R=1$ bit/s/Hz. $L=2$ ms. (a) Connection-free and connection-based random access with CB. $n\in\{20,50,100,500\}$. $\Delta_{S,P}=\Delta_{F,P}=1$ ms. $\Delta_{S,N}=3$ ms. $\Delta_{F,N}=1.5$ ms. (b) Connection-free random access with CB and BEB. $n=50$. $\Delta_{S,P}=\Delta_{F,P}=1$ ms. $K\in\{1,2,3,4\}$.}
    \label{fig5-1}
    \vspace{-0.4cm}
\end{figure*}

It can be seen from (\ref{eq2-8}) that to derive the delay-optimal sensing bound $\tilde{\sigma}_C^{\ast}$, the minimum mean queueing delay of data packets in the unit of ms $\tilde{T}_{\min}^A$ with Aloha and $\tilde{T}_{\min}^C$ with CSMA need to be compared. Note that $\tilde{T}_{\min}=\sigma\cdot\overline{T}_{\min}$, where the minimum mean queueing delay of data packets $\overline{T}_{\min}$ in the unit of time slots has been obtained in Section \ref{sec4}. It was shown that both $\overline{T}_{\min}^A$ with Aloha and $\overline{T}_{\min}^C$ with CSMA closely depend on the aggregate packet input rate $\hat{\lambda}$, which can further be written from (\ref{eq2-1}) as $\hat{\lambda}=\tfrac{\tilde{\lambda}\sigma}{RL}$, where $\tilde{\lambda}=n\lambda_b$ denotes the aggregate bit input rate. For a fair comparison of $\tilde{T}_{\min}^A$ and $\tilde{T}_{\min}^C$, assume that the aggregate bit input rates $\tilde{\lambda}$ for both Aloha and CSMA are equal, with which the aggregate packet input rates with Aloha and CSMA should satisfy $\tfrac{\hat{\lambda}_A}{\sigma_A}=\tfrac{\hat{\lambda}_C}{\sigma_C}$.

The delay-optimal sensing bound $\tilde{\sigma}_C^{\ast}$ for the given aggregate bit input rate $\tilde{\lambda}$ can then be obtained by combining (\ref{eq2-8}) with (\ref{eq2-1}), (\ref{eq2-2-1}), (\ref{eq2-7}), (\ref{eq3-1-1-1.5}), (\ref{eq3-1-2-1.25})-(\ref{eq3-1-2-1.5}), (\ref{eq3-2-1-7})-(\ref{eq3-2-1-8}), (\ref{eq3-2-2-5}), (\ref{eq3-2-2-7}), (\ref{eq4-1-1})-(\ref{eq4-1-1-1}), (\ref{eq4-1-1-4})-(\ref{eq4-2-1}), which is illustrated in Fig. \ref{fig5-1}. Note that each node's queue is saturated with unbounded mean queuing delay if $\tilde{\lambda}\geq\tilde{\lambda}_{\max}$, where the maximum data throughput in the unit of bit/s/Hz $\tilde{\lambda}_{\max}$ can be obtained by combining (\ref{eq2-3.5}) with (\ref{eq3-2-2}) for Aloha and with (\ref{eq3-2-3}) for CSMA. In the saturated condition, the optimal throughput performance becomes the main concern. The throughput-optimal sensing bound $\sigma_C^{\ast}$, given by (\ref{eq2-5})-(\ref{eq2-6}), is also plotted in Fig. \ref{fig5-1}.

Fig. \ref{fig5-1-1} presents the delay-optimal sensing bound $\tilde{\sigma}_C^{\ast}$ and throughput-optimal sensing bound $\sigma_C^{\ast}$ with connection-free and connection-based random access. It can be seen that with the given overhead parameters, $\tilde{\sigma}_C^{\ast,P}$ and $\sigma_C^{\ast,P}$ in the connection-free case are larger than $\tilde{\sigma}_C^{\ast,N}$ and $\sigma_C^{\ast,N}$ in the connection-based case, respectively. Note that a larger sensing bound indicates that sensing is more likely beneficial. Intuitively, as the time of each failed transmission $L+\Delta_{F,P}$ with connection-free access is typically larger than $\Delta_{F,N}$ with connection-based access, the benefit from sensing in reducing the collision probability is more significant in the connection-free case. As a result, the network with connection-free random access is more likely to benefit from sensing than that with connection-based random access.

Fig. \ref{fig5-1-1} also illustrates how $\tilde{\sigma}_C^{\ast}$ and $\sigma_C^{\ast}$ vary with the aggregate bit input rate $\tilde{\lambda}$ under different settings of the number of nodes $n$. It is clear that different from the throughput-optimal sensing bound $\sigma_C^{\ast}$, the delay-optimal sensing bound $\tilde{\sigma}_C^{\ast}$ is closely determined by $\tilde{\lambda}$ and $n$. As the number of nodes $n$ grows, $\tilde{\sigma}_C^{\ast}$ would be increased. It suggests that when a massive number of nodes contend for channel access, the use of sensing is very likely to improve the optimal delay performance.

In addition to the aggregate bit input rate $\tilde{\lambda}$ and the number of nodes $n$, the delay-optimal sensing bound $\tilde{\sigma}_C^{\ast}$ further depends on the backoff scheme. $\tilde{\sigma}_C^{\ast,CB}$ with CB and $\tilde{\sigma}_C^{\ast,BEB}$ with BEB are illustrated in Fig. \ref{fig5-1-2}. It can be observed that the delay-optimal sensing bound $\tilde{\sigma}_C^{\ast,BEB}$ with BEB is lower than $\tilde{\sigma}_C^{\ast,CB}$ with CB, and is decreased as the cutoff phase $K$ increases from 1 to 4. Intuitively, with BEB and properly selected $K$, the contention among nodes can be greatly alleviated, in which case the benefit from sensing becomes insignificant, and the sensing time required for CSMA to outperform Aloha would be lower. This implies that a network with CB is more likely to benefit from sensing than that with BEB.

\subsection{Case Study: Random Access Schemes in 5G Networks}\label{sec5-2}

To demonstrate how to apply the above analysis to practical networks, let us consider 5G cellular networks operating in the licensed spectrum. Specifically, we focus on the scenario where a massive number of MTDs transmit their short packets to the gNB through the 2-step RA-SDT and 4-step RA-SDT schemes \cite{3gpp.38.321}.

\begin{figure*}[t]
    \centering
    \subfloat[]
    {
		\includegraphics[width=0.75\textwidth, height=0.073\textwidth]{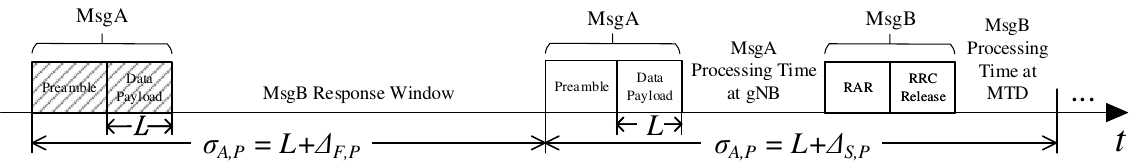}
        \label{fig5-2-1}
    }
    \vspace{-0.1cm}
    \subfloat[]
    {
        \includegraphics[width=0.75\textwidth, height=0.073\textwidth]{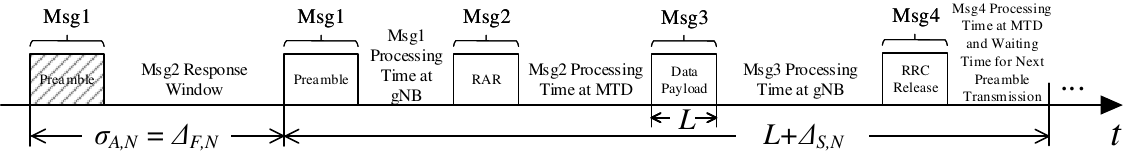}
        \label{fig5-2-2}
    }
    \caption{Graphic illustration of (a) 2-step RA-SDT and (b) 4-step RA-SDT schemes in 5G cellular networks \cite{3gpp.38.321}.}
    \label{fig5-2}
    \vspace{-0.2cm}
\end{figure*}

As Fig. \ref{fig5-2-1} illustrates, with 2-step RA-SDT, each MTD that has data packets would start its MsgA transmission, including a preamble and its data payload, at the beginning of each available Physical Random Access CHannel (PRACH) occasion with a certain probability\footnote{Note that in 5G networks, the transmission probability of each MTD is determined by the Access Class Barring (ACB) factor $q_{ACB}$ and the Uniform Backoff (UB) window size $W_{UB}$ (in the unit of ms) \cite{3gpp.38.331}. Let $\tilde{W}_{UB}$ denote the UB window size in the unit of time slots, which can be obtained from $W_{UB}$ as $\tilde{W}_{UB}=\left\lfloor \tfrac{W_{UB}}{\sigma}\right\rfloor + 1$, with $\sigma$ denoting the time slot length. Then the transmission probability is given by $\tfrac{2 q_{ACB}}{\tilde{W}_{UB}+1}$.}. If the Random Access Response (RAR) message and the Radio Resource Control (RRC) Release message are received within the MsgB Response Window, then the MTD knows that its MsgA transmission has been successful. With 4-step RA-SDT, on the other hand, each MTD first transmits a preamble in Msg1 with a certain probability at each available PRACH occasion, and transmits its data payload\footnote{Note that a control message, RRC Resume Request, is also transmitted in MsgA and Msg3. It is not included in the overhead time shown in Fig. \ref{fig5-2} because RRC Resume Request and data payload can be transmitted in different frequency-domain resource blocks \cite{3gpp.38.133, 3gpp.38.213, 3gpp.38.214}, and no extra overhead in the time domain is brought by this message.} in Msg3 only after receiving the RAR message within the Msg2 Response Window, as shown in Fig. \ref{fig5-2-2}. Note that for the 5G networks operating at the licensed spectrum, MTDs do not perform carrier sensing before MsgA or Msg1 transmissions and do not change their transmission probabilities according to the number of collisions. Therefore, 2-step RA-SDT and 4-step RA-SDT are indeed connection-free (grant-free) and connection-based (grant-based) Aloha protocols with CB, respectively.

Without loss of generality, we consider one single preamble.\footnote{Note that in the 5G system, multiple preambles can be selected by MTDs. A collision occurs if more than one MTD selects the same preamble and transmits concurrently. As the main focus of this work is on the comparison of sensing-free and sensing-based random access, for the sake of discussion, we only consider the case where one preamble is available.} The time slot length is set to the interval between two consecutive PRACH occasions available for MTDs.\footnote{As \cite{3gpp.38.213} specifies, PRACH occasions for each MTD's preamble transmission are available periodically, that is, each MTD can start its MsgA or Msg1 transmission only at the beginning of each periodic PRACH occasion.} Table \ref{table1} lists the typical values of key system parameters \cite{IMT-2020, 3gpp.38.213, 3gpp.38.214, 3gpp.38.133}. With the Transmission Time Interval (TTI) set to $0.5$ ms, the data payload transmission time $L$, the overhead time $\Delta_S$ for each successful transmission and $\Delta_F$ for each failed transmission in 2-step RA-SDT and 4-step RA-SDT can be obtained as
\begin{equation}
    \begin{split}
        & L=0.5\,{\rm{ms,}}\;\Delta_{S,P} = \Delta_{F,P} = 5.5\,{\rm{ms}}, \\
        & \Delta_{S, N} = 7.5\,{\rm{ms}},\;\Delta_{F, N} = 2\,{\rm{ms}},
    \end{split}
    \label{eq5-2-1}
\end{equation}
according to Fig. \ref{fig5-2} and Table \ref{table1}.

\begin{table}[t]
    \scriptsize
    \centering
    \caption{System Parameter Setting \cite{IMT-2020, 3gpp.38.213, 3gpp.38.214, 3gpp.38.133}}
    \vspace{-0.1cm}
    \label{table1}
    \resizebox{\columnwidth}{!}{
    \begin{tabular}{|c|c|c|}
    \hline
                                    & Parameter                                                          & Value            \\ \hline
    \multirow{3}{*}{\begin{tabular}[c]{@{}c@{}}Grant-Free \\2-Step RA-SDT\end{tabular}}  & MsgA Processing Time at gNB & 1 TTI + 3 ms \\
                                    & MsgB Processing Time at MTD                                        & 1 TTI            \\
                                    & MsgB Response Window Length                                              & 5 ms   \\ \hline
    \multirow{6}{*}{\begin{tabular}[c]{@{}c@{}}Grant-Based \\4-Step RA-SDT\end{tabular}} & Msg1 Processing Time at gNB & 1 TTI              \\
                                    & Msg2 Processing Time at MTD                                        & 3 TTIs            \\
                                    & Msg3 Processing Time at gNB                                        & 3 ms   \\
                                    & Msg4 Processing Time at MTD                                        & 1 TTI            \\
                                    & Waiting Time for Next Preamble Transmission                               & 0.5 ms \\
                                    & Msg2 Response Window Length                                              & 1.5 ms \\ \hline
    \multirow{3}{*}{Common Setting} & Preamble/RAR/RRC Release Transmission Time      & 1 TTI            \\
    & Data Payload Transmission Time $L$         & 1 TTI            \\
    & Data Encoding Rate $R$                                      & 0.3066 bit/s/Hz         \\ \hline
    \end{tabular}
    }
\end{table}

\begin{figure}[t]
    \vspace{-0.05cm}
    \centering
    \includegraphics[width=0.39\textwidth, height=0.22\textwidth]{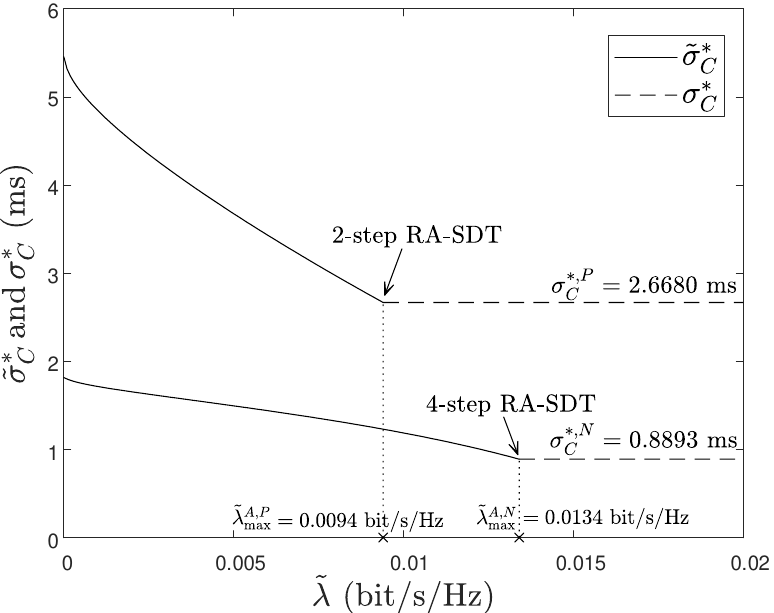}
    \caption{Delay-optimal sensing bound $\tilde{\sigma}_C^{\ast}$ and throughput-optimal sensing bound $\sigma_C^{\ast}$ with grant-free 2-step RA-SDT and grant-based 4-step RA-SDT versus aggregate bit input rate $\tilde{\lambda}$. $n=500$.}
    \label{fig5-3}
\end{figure}

To see whether sensing can improve the performance of the 2-step RA-SDT and 4-step RA-SDT schemes, Fig. \ref{fig5-3} illustrates the delay-optimal sensing bound $\tilde{\sigma}_C^{\ast}$ and throughput-optimal sensing bound $\sigma_C^{\ast}$. It can be seen that the delay-optimal sensing bound $\tilde{\sigma}_C^{\ast}$ crucially depends on the aggregate bit input rate $\tilde{\lambda}$, and is larger than the throughput optimal sensing bound $\sigma_C^{\ast}$, which is equal to 2.6680 ms with 2-step RA-SDT and 0.8893 ms with 4-step RA-SDT. This indicates that for 2-step RA-SDT and 4-step RA-SDT, if sensing is incorporated with the sensing time $\sigma_C$ lower than 2.6680 ms and 0.8893 ms, respectively, both the optimal delay performance and optimal throughput performance would be improved.

Note that for 5G networks operating at the unlicensed spectrum, sensing is required to be performed before each message transmission in order not to disrupt transmissions from other coexisting networks such as WiFi \cite{3gpp.37.213}. For 5G networks in the licensed spectrum, although there is no need for nodes to sense transmissions from other networks, the analysis above implies that sensing is very likely to bring gains in the delay performance thanks to alleviated collisions among nodes within the network. To take a closer look at how much gain can be brought by sensing, we further propose sensing-based 2-step RA-SDT and 4-step RA-SDT schemes for 5G licensed spectrum,\footnote{Note that different from the existing sensing-based access schemes of 5G networks in the unlicensed spectrum, where MTDs and gNB should sense before each step of the random access procedure to ensure that the channel is not occupied by other systems, here only MTDs need to sense before the MsgA or Msg1 transmission. That is because in the licensed spectrum, if MsgA or Msg1 is successful, the following message transmissions would not be interrupted.} where the sensing time $\sigma_C$ is designed to be 1 TTI (i.e., 0.5 ms), and compare the performance of the sensing-free RA-SDT schemes with the proposed sensing-based RA-SDT schemes. Specifically, we propose the following system setting: \romannumeral1) PRACH occasions are available per TTI, with each lasting for 1 TTI; \romannumeral2) MTDs have the sensing capability; \romannumeral3) MTDs know the time of each successful transmission (i.e., $L+\Delta_{S,P}$ or $L+\Delta_{S,N}$) and the time of each failed transmission (i.e., $L+\Delta_{F,P}$ or $\Delta_{F,N}$), which can be broadcast by gNB. The basic idea is that each MTD needs to sense the PRACH occasion for 1 TTI before each transmission. If the PRACH occasion is sensed idle, the MTD would perform 2-step or 4-step RA-SDT procedure; otherwise, the MTD would mute itself until the ongoing RA-SDT procedure is completed. The time of mute period is determined by the time of each successful or failed transmission. The details of the proposed sensing-based 2-step RA-SDT and 4-step RA-SDT are presented in Algorithm \ref{alg1} and Algorithm \ref{alg2}, respectively.

\newcounter{Stepnum}
\setcounter{Stepnum}{1}
\algnewcommand\algorithmicinput{\textbf{Step}}
\algnewcommand\StepA{\item[\algorithmicinput] \textbf{1: }}
\algnewcommand\StepB{\item[\algorithmicinput] \textbf{2: }}
\algnewcommand\StepC{\item[\algorithmicinput] \textbf{3: }}
\algnewcommand\StepD{\item[\algorithmicinput] \textbf{4: }}

\begin{algorithm}[t]
    \footnotesize
    \caption{Sensing-Based 2-Step RA-SDT}
    \label{alg1}
    \begin{algorithmic}[1]
        \StepA Sense the PRACH occasion for 1 TTI. If the PRACH occasion is sensed idle, perform Step 2 in the next PRACH occasion; otherwise, go to step 4.
        \StepB Transmit MsgA with a certain probability, and go to Step 3. If MsgA transmission is not performed, repeat Step 1-2.
        \StepC Start the MsgB Response Window after sending MsgA. If MsgB is correctly received within the MsgB Response Window, the preceding MsgA transmission is considered to be successful. Otherwise, go back to Step 1.
        \StepD Mute for $L+\Delta_{S,P}$ ms from the beginning of the PRACH occasion that was sensed busy. Then go back to Step 1.
    \end{algorithmic}
\end{algorithm}

\begin{algorithm}[t]
    \footnotesize
    \caption{Sensing-Based 4-Step RA-SDT}
    \label{alg2}
    \begin{algorithmic}[1]
        \StepA Sense the PRACH occasion for 1 TTI. If the PRACH occasion is sensed idle, perform Step 2 in the next PRACH occasion; otherwise, go to step 4.
        \StepB Transmit Msg1 with a certain probability, and go to Step 3. If Msg1 transmission is not performed, repeat Step 1-2.
        \StepC Start the Msg2 Response Window after sending Msg1. If Msg2 is correctly received within the Msg2 Response Window, the preceding Msg1 transmission is considered to be successful, and Msg3 will be transmitted over the allocated uplink resources. Otherwise, go back to Step 1.
        \StepD Mute from the beginning of the PRACH occasion that was sensed busy. If a Msg2 indicating that another MTD has obtained the uplink grant is received, mute for $L+\Delta_{S,N}$ ms; otherwise, mute for $\Delta_{F,N}$ ms. Then go back to Step 1.
    \end{algorithmic}
\end{algorithm}

\begin{figure*}[t]
    \centering
    \subfloat[]{
        \includegraphics[width=0.39\textwidth, height=0.23\textwidth]{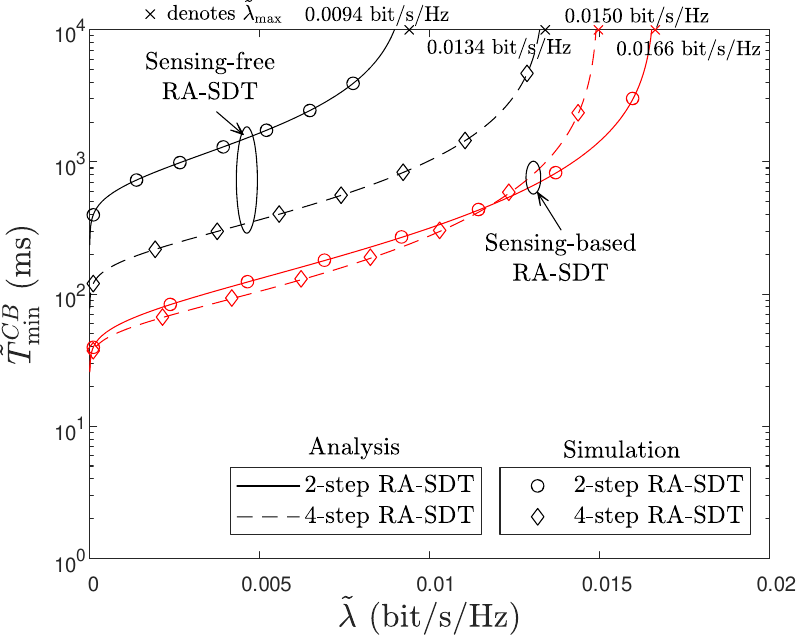}
        \label{fig5-4-1}
    }\hspace{1.2cm}
    \subfloat[]{
        \includegraphics[width=0.39\textwidth, height=0.23\textwidth]{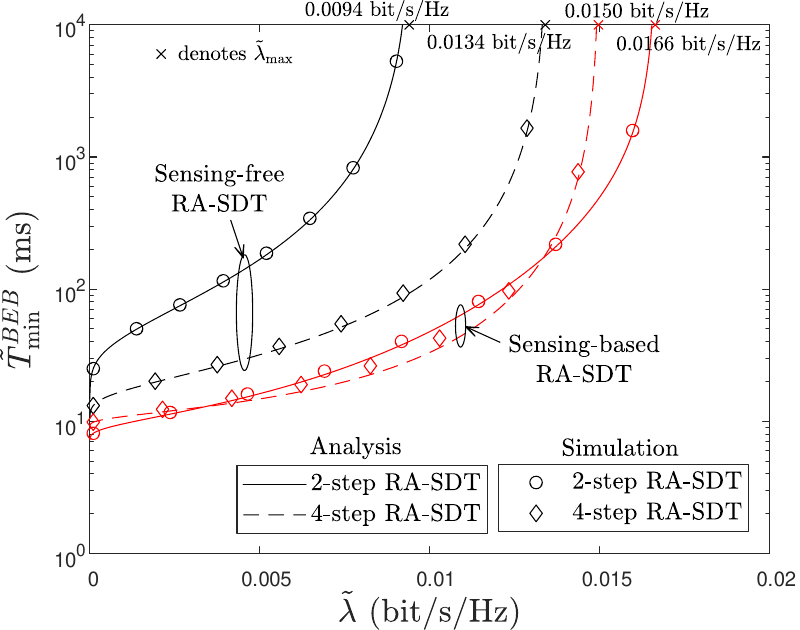}
        \label{fig5-4-2}
    }
    \caption{Analytical and simulated minimum mean queueing delay of data packets in the unit of ms $\tilde{T}_{\min}^{A}$ with sensing-free RA-SDT schemes and $\tilde{T}_{\min}^{C}$ with proposed sensing-based RA-SDT schemes versus aggregate bit input rate $\tilde{\lambda}$. $n=500$. (a) With CB. (b) With BEB. $K=4$.}
    \label{fig5-4}
    \vspace{-0.3cm}
\end{figure*}

The proposed sensing-based 2-step RA-SDT and 4-step RA-SDT schemes are essentially grant-free CSMA and grant-based CSMA, respectively, with the sensing time $\sigma_C=0.5$ ms. Given the preceding parameter setting, the minimum mean queueing delay of data packets in the unit of ms $\tilde{T}_{\min}^{C,P}$ with sensing-based 2-step RA-SDT and $\tilde{T}_{\min}^{C,N}$ with sensing-based 4-step RA-SDT can be obtained by substituting (\ref{eq5-2-1}) and $\sigma_C=0.5$ ms into (\ref{eq2-7}), (\ref{eq3-1-2-1.25})-(\ref{eq3-1-2-1.5}), (\ref{eq3-2-1-8}), (\ref{eq3-2-2-7}), (\ref{eq4-1-1})-(\ref{eq4-1-1-1}), (\ref{eq4-1-1-6})-(\ref{eq4-2-1}).

Fig. \ref{fig5-4-1} presents the optimal delay performance of both the proposed sensing-based RA-SDT schemes and the sensing-free RA-SDT schemes adopted in current 5G networks. It can be seen that compared to the sensing-free RA-SDT schemes, the proposed sensing-based RA-SDT schemes achieve a lower minimum mean queueing delay of data packets in addition to higher maximum data throughput. Besides, different from the existing sensing-free case, where grant-based 4-step RA-SDT outperforms grant-free 2-step RA-SDT in terms of the optimal delay performance, with sensing incorporated, no significant gains are brought by 4-step RA-SDT. It corroborates the analysis in Section \ref{sec5-1} that a network with connection-free random access usually benefits more from sensing than that with connection-based random access.

Note that the backoff scheme adopted by the existing RA-SDT mechanism is CB. To investigate how the backoff scheme affects the delay performance of the existing and proposed RA-SDT mechanism, we replace CB with BEB, and illustrate the minimum mean queueing delay of data packets in Fig. \ref{fig5-4-2}. It can be observed that when BEB is adopted, the proposed sensing-based RA-SDT schemes also outperform the existing sensing-free RA-SDT schemes in terms of the optimal delay performance, though the gain is not as significant as that with CB. Furthermore, we can see from Fig. \ref{fig5-4-1} and Fig. \ref{fig5-4-2} that the minimum mean queueing delay is significantly reduced by replacing CB with BEB. The analysis above suggests that incorporating sensing and BEB into current RA-SDT schemes could be highly desirable for improving the delay performance.

\section{Conclusion}\label{sec6}

In this paper, the mean queueing delay of data packets with sensing-free Aloha and sensing-based CSMA is characterized and optimized, based on which the criteria for beneficial sensing in terms of the optimal delay performance are further obtained. Specifically, for each node's queue, by establishing and solving the fixed-point equations of steady-state probability of successful transmission of HOL packets based on a unified HOL-packet model, the minimum mean queueing delay of data packets with Bernoulli arrivals and the corresponding optimal initial transmission probability are derived and shown to be closely determined by the data input rate of each node, the number of nodes, and the design features including sensing-free or sensing-based, connection-free or connection-based, and backoff. Based on the minimum mean queueing delay derived for sensing-free Aloha and sensing-based CSMA, the delay-optimal sensing bound, below which CSMA is superior to Aloha in terms of the optimal delay performance, is further characterized and shown to crucially depend on the data input rates of nodes and the backoff scheme. It is observed that the delay-optimal sensing bound with CB is larger than that with BEB, indicating that a network with CB is more likely to benefit from sensing compared to that with BEB.

The analysis is further applied to the grant-free 2-step and grant-based 4-step RA-SDT schemes of 5G networks in the licensed spectrum to demonstrate when and how significant their delay performance can be improved by the use of sensing. It is shown that the minimum mean queueing delay with both grant-free 2-step RA-SDT and grant-based 4-step RA-SDT can be substantially reduced by incorporating sensing into the access procedure, and the 2-step RA-SDT benefits more. Moreover, the optimal delay performance can further be improved by replacing the currently adopted CB with BEB, though the gain brought by sensing is not as significant as that in the CB case. The analysis suggests that introducing sensing and BEB into RA-SDT schemes could be beneficial for facilitating low-latency communications of MTDs.

\appendices

\section{Derivation of (\ref{eq3-2-1-2})-(\ref{eq3-2-1-6})}\label{sec_ap1}

In (\ref{eq3-2-1-1}), the probability that the node intends to request can be written as
\begin{equation}
    \Pr\{{\rm The\,node\,intends\,to\,request}\} = \rho\sum_{u\in\mathcal{S}}\varphi_u,
    \label{eq-ap1-1}
\end{equation}
where $\rho$ denotes the probability of the node's queue being busy, which is given by
\begin{equation}
    \rho =
    \begin{cases}
        \tfrac{\lambda}{\mu} & \lambda<\mu \\
        1 & \lambda\geq\mu,
    \end{cases}
    \label{eq-ap1-2}
\end{equation}
and $\varphi_u$ denotes the probability that the node has a State-$u$ HOL packet and intends to request given that its queue is busy, $u\in\mathcal{S}$. On the other hand, by noting that when the channel is accessible, the node either has an empty queue, or has a busy queue and intends to request, the probability that the node intends to request given that the channel is accessible can be written as
\begin{equation}
    \begin{split}
        & \Pr\{{\rm The\,node\,intends\,to\,request}\,|\,{\rm Channel\,is\,accessible}\} \\
        = & \tfrac{\rho\sum_{u\in\mathcal{S}}\varphi_u}{1-\rho+\rho\sum_{u\in\mathcal{S}}\varphi_u}.
    \end{split}
    \label{eq-ap1-3}
\end{equation}
By combining (\ref{eq3-2-1-1}) with (\ref{eq-ap1-1})-(\ref{eq-ap1-3}), the conditional steady-state probability of the channel being accessible given that the node intends to request, $\tilde{\alpha}$, can be obtained as (\ref{eq3-2-1-2}).

In (\ref{eq3-2-1-2}), the probability that the node has a State-$u$ HOL packet and intends to request given that its queue is busy, $\varphi_u$, depends on which type of random access, Aloha or CSMA, is adopted. Specifically, with Aloha, each node intends to request only when its HOL packet is at State B$_k$ or the first time slot of State T; with CSMA, each node intends to request only when its HOL packet is at State R$_k$, $k\in\{0,\cdots,K\}$. Therefore, $\varphi_u$ can be written as (\ref{eq3-2-1-3}) and (\ref{eq3-2-1-4}) for Aloha and CSMA, respectively. On the other hand, the unconditional probability of the channel being accessible to all the nodes $\alpha$ is also dependent on sensing-free or sensing-based, and will be derived for sensing-free Aloha and sensing-based CSMA in the following subsections, respectively.

\subsubsection{Derivation of $\alpha^A$}\label{sec_ap1_1}

With Aloha, the channel is not accessible to all the nodes if it is reserved by data transmission. Specifically, the channel would be reserved for $\tau_T^A-1$ time slots if there is a successful request. We then have
\begin{equation}
    (\tau_T^A-1)\cdot \Pr\{{\rm There\,is\,a\,successful\,request}\} = 1 - \alpha^A.
    \label{eq-ap1-1-1}
\end{equation}
With the collision model, given that the channel is accessible to all the nodes, the probability that there is a successful request can be written as $n\omega(1-\omega)^{n-1}$, where $\omega$ denotes the probability that the node has a busy queue and makes a request given that the channel is accessible. Therefore, the probability that there is a successful request can be written as
\begin{equation}
    \Pr\{{\rm There\,is\,a\,successful\,request}\} = n\omega(1-\omega)^{n-1}\cdot \alpha.
    \label{eq-ap1-1-2}
\end{equation}
Further note that with the collision model, the steady-state probability of successful transmission of HOL packets given that the channel is accessible $p$ can be expressed as $p=(1-\omega)^{n-1}$. By combining $p=(1-\omega)^{n-1}$ with (\ref{eq-ap1-1-1})-(\ref{eq-ap1-1-2}), $\alpha^A$ can be obtained as (\ref{eq3-2-1-5}).

\subsubsection{Derivation of $\alpha^C$}\label{sec_ap1_2}

With CSMA, the channel is not accessible to all the nodes if it is sensed busy. Specifically, the channel would be sensed busy for $\tau_T^C$ time slots if there is a successful request, or $\tau_F^C$ time slots if there is a failed request. We then have
\begin{equation}
    \begin{split}
        & \tau_T^C\cdot \Pr\{{\rm There\,is\,a\,successful\,request}\} + \tau_F^C\cdot \Pr\{{\rm There\,is\,a} \\
        & {\rm \,failed\,request}\} = 1 - \alpha^C.
    \end{split}
    \label{eq-ap1-2-1}
\end{equation}
When the collision model is adopted, the probability that there is a successful request is given by (\ref{eq-ap1-1-2}), and the probability that there is a failed request can be written as
\begin{equation}
    \Pr\{{\rm There\,is\,a\,failed\,request}\} \!=\! \big(1-(1-\omega)^n-n\omega(1-\omega)^{n-1}\big)\cdot \alpha,
    \label{eq-ap1-2-2}
\end{equation}
where $\omega$ denotes the probability that the node has a busy queue and makes a request given that the channel is accessible. By noting that $p=(1-\omega)^{n-1}$ and combining (\ref{eq-ap1-1-2})-(\ref{eq-ap1-2-2}), $\alpha^C$ can be obtained as (\ref{eq3-2-1-6}).

\section{Derivation of (\ref{eq3-2-2-2})}\label{sec_ap2}

The probability that the node has a busy queue and makes a request given that the channel is accessible, $\omega$, can be written as
\begin{equation}
    \begin{split}
        \omega = & \Pr\{{\rm The\,node\,makes\,a\,request}\,|\,{\rm The\,node's\,queue\,} \\
        & {\rm is\,busy\,and\,channel\,is\,accessible}\}\cdot\Pr\{{\rm The\,node's\,} \\
        & {\rm queue\,is\,busy\,}\,|\,{\rm Channel\,is\,accessible}\}.
    \end{split}
\label{eq-ap2-1}
\end{equation}

Let $\omega^{(1)}$ denote the first item on the right-hand side of (\ref{eq-ap2-1}). For each node, given that its queue is busy and the channel is accessible, it must have a HOL packet and intend to request. $\omega^{(1)}$ can then be written as
\begin{equation}
    \omega^{(1)} = \tfrac{\sum_{u\in\mathcal{S}}\varphi_u q_u}{\sum_{u\in\mathcal{S}}\varphi_u},
\label{eq-ap2-2}
\end{equation}
where $\varphi_u$ is the probability that the node has a State-$u$ HOL packet and intends to request given that its queue is busy, and $q_u$ denotes the transmission probability of the node given that it has a State-$u$ HOL packet and intends to request. $\varphi_u$ is given by (\ref{eq3-2-1-3}) and (\ref{eq3-2-1-4}) for Aloha and CSMA, respectively, and $q_u$ is given by
\begin{equation}
    q_u^A =
    \begin{cases}
        q_0\mathcal{Q}(k) & u\in\{{\rm B}_k\}_{k=0,\cdots,K} \\
        q_0 & u\in\{{\rm T}\},
    \end{cases}
    \label{eq-ap2-3}
\end{equation}
and
\begin{equation}
    q_u^C =
    \begin{cases}
        q_0\mathcal{Q}(k) & u\in\{{\rm R}_k\}_{k=0,\cdots,K} \\
        0 & u\in\{{\rm T},{\rm F}_0,\cdots,{\rm F}_K\},
    \end{cases}
    \label{eq-ap2-4}
\end{equation}
for Aloha and CSMA, respectively.

Let $\omega^{(2)}$ denote the second item on the right-hand side of (\ref{eq-ap2-1}), which can be written as
\begin{equation}
    \omega^{(2)} = \tfrac{\rho}{\alpha}\Pr\{{\rm Channel\,is\,accessible}\,|\,{\rm The\,node's\,queue\,is\,busy}\},
\label{eq-ap2-5}
\end{equation}
where $\rho$ is the probability that the node's queue is busy and given by (\ref{eq-ap1-2}), and $\alpha$ is the unconditional probability that the channel is accessible to all the nodes. Further note that
\begin{equation}
    \begin{split}
        & \Pr\{{\rm Channel\,is\,accessible}\,|\,{\rm The\,node's\,queue\,is\,busy}\} \\
        = & \sum_{u\in\mathcal{S}} \Pr\{{\rm Channel\,is\,accessible}\,|\,{\rm The\,node\,has\,a\,State-}u \\
        & {\rm HOL\,packet\,and\,intends\,to\,request}\}\cdot \Pr\{{\rm The\,node\,has\,} \\
        & {\rm a\,State-}u\,{\rm HOL\,packet\,and\,intends\,to\,request\,}|\,{\rm The} \\
        & {\rm node's\,queue\,is\,busy}\} \\
        = & \sum_{u\in\mathcal{S}}\tilde{\alpha}\varphi_u.
    \end{split}
\label{eq-ap2-5.5}
\end{equation}
By combining (\ref{eq-ap2-5.5}) with (\ref{eq-ap2-5}), $\omega^{(2)}$ can be further written as
\begin{equation}
    \omega^{(2)} = \tfrac{\rho\tilde{\alpha}}{\alpha}\sum_{u\in\mathcal{S}}\varphi_u,
\label{eq-ap2-6}
\end{equation}

By combining (\ref{eq-ap2-2}) and (\ref{eq-ap2-6}), (\ref{eq-ap2-1}) can be written as
\begin{equation}
    \omega = \tfrac{\rho\tilde{\alpha}}{\alpha}\sum_{u\in\mathcal{S}}\varphi_u q_u.
\label{eq-ap2-7}
\end{equation}
Note that for both Aloha and CSMA, we have $\sum_{u\in\mathcal{S}}\varphi_u q_u = \tfrac{\tilde{\pi}_T}{p\tilde{\alpha}\tau_T}$ according to (\ref{eq3-1-1}), (\ref{eq3-1-1-1}), (\ref{eq3-1-1-3}), (\ref{eq3-2-1-3}), (\ref{eq-ap2-3}) and (\ref{eq3-1-2-1}), (\ref{eq3-1-2-3}), (\ref{eq3-2-1-4}), (\ref{eq-ap2-4}). By combining $\sum_{u\in\mathcal{S}}\varphi_u q_u = \tfrac{\tilde{\pi}_T}{p\tilde{\alpha}\tau_T}$, (\ref{eq3-1-2}), (\ref{eq-ap1-2}) with (\ref{eq-ap2-7}), the probability that the node has a busy queue and makes a request given that the channel is accessible, $\omega$, can be obtained as (\ref{eq3-2-2-2}).

\section{Derivation of (\ref{eq4-1-1-2})-(\ref{eq4-1-1-3})}\label{sec_ap3}

Note that each HOL packet is initially at State T, and its service time is equal to the number of time slots from its first state transition till its service completion (i.e., after it shifts back to State T and stays there for $\tau_T$ time slots). The service time distribution can then be written as
\begin{equation}
    \Pr\{D=i\} = \sum_{u\in\mathcal{S}} P_{T,u}\Pr\{D_u=i\},
\label{eq-ap3-1}
\end{equation}
where $P_{T,u}$ denotes the transition probability from State T to State $u$, and $D_u$ denotes the number of time slots from the HOL packet shifting to State $u$ till the service completion, $u\in\mathcal{S}$. The probability generating function of service time of each node's queue can then be obtained as (\ref{eq4-1-1-2}) by noting that $G_D(z)=\sum_{i=0}^{\infty}\Pr\{D=i\}z^i$ and $G_{D_u}(z)=\sum_{i=0}^{\infty}\Pr\{D_u=i\}z^i$.

The time from a HOL packet shifting to State T till the service completion $D_T$ is equal to the holding time at State T $Y_T$, which is deterministic. Therefore, the probability distribution of $D_T$ can be written as 
\begin{equation}
    \Pr\{D_T=i\} =
    \begin{cases}
        1 & i = \tau_T \\
        0 & {\rm otherwise}.
    \end{cases}
\label{eq-ap3-2}
\end{equation}
On the other hand, for $u\in\mathcal{S}\backslash \{{\rm{T}}\}$, the probability distribution of $D_u$ can be written as
\begin{equation}
    \Pr\{D_u=i\} = \sum_{v\in\mathcal{S}} P_{u,v}\Pr\{Y_u+D_v=i\},
\label{eq-ap3-3}
\end{equation}
where $P_{u,v}$ is the transition probability from State $u$ to State $v$, $Y_u$ is the holding time at State $u$, and $D_v$ is the time from the beginning of State $v$ till the service completion. By combining (\ref{eq-ap3-2})-(\ref{eq-ap3-3}) and noting that $Y_u$ and $D_v$ are independent with each other, the probability generating functions of $\{D_u\}_{u\in\mathcal{S}}$ can then be obtained as (\ref{eq4-1-1-3}).

\balance

\normalem
\bibliographystyle{IEEEtran}
\bibliography{IEEEabrv, ref}

\end{document}